\documentclass[aps,superscriptaddress,showpacs,twocolumn,dvips]{revtex4}
\usepackage{feynmp}
\usepackage{amssymb}
\usepackage{epsfig}
\usepackage{graphicx}
\usepackage{subfigure}
\usepackage{mathrsfs}
\usepackage{hyperref}
\usepackage{cases}
\usepackage{bbm}
\usepackage{CJK}
\usepackage{xcolor}

\begin{document}

\begin{CJK*}{GBK}{song}


\title{Cooper instability generated by attractive fermion-fermion interaction in the two-dimensional
semi-Dirac semimetals}

\date{\today}

\author{Yao-Ming Dong}
\affiliation{Department of Physics, Tianjin University, Tianjin 300072, P.R. China}

\author{Dong-Xing Zheng}
\affiliation{Department of Physics, Tianjin University, Tianjin 300072, P.R. China}

\author{Jing Wang}
\altaffiliation{Corresponding author: jing$\textunderscore$wang@tju.edu.cn}
\affiliation{Department of Physics, Tianjin University, Tianjin 300072, P.R. China}
\affiliation{Department of Modern Physics, University of Science and
Technology of China, Hefei, Anhui 230026, P.R. China}

\begin{abstract}
Cooper instability associated with superconductivity in the two-dimensional
semi-Dirac semimetals is attentively studied in the presence of attractive Cooper-pairing
interaction, which is the projection of an attractive fermion-fermion interaction. Performing the
standard renormalization group analysis shows that the Cooper theorem is violated at
zero chemical potential but instead Cooper instability can be generated only if
the absolute strength of fermion-fermion coupling exceeds certain
critical value and transfer momentum is restricted to a confined region, which
is determined by the initial conditions.
Rather, the Cooper theorem would be instantly restored once a finite chemical potential
is introduced and thus a chemical potential-tuned phase transition is expected.
Additionally, we briefly examine the effects of impurity scatterings on the Cooper instability
at zero chemical potential, which in principle are harmful to Cooper instability although
they can enhance the density of states of systems. Furthermore, the influence of competition between
a finite chemical potential and impurities upon the Cooper instability is also simply investigated.
These results are expected to provide instructive clues for exploring unconventional superconductors
in the kinds of semimetals.
\end{abstract}


\maketitle



\section{Introduction}

Accompanying with the remarkable developments in the Dirac fermions~\cite{Novoselov2005Nature,Neto2009RMP,Kane2007PRL,Roy2009PRB,Moore2010Nature,Hasan2010RMP,
Qi2011RMP,Sheng2012Book,Bernevig2013Book,Burkov2011PRL,Yang2011PRB,Savrasov2011PRB,
Huang2015PRX,Weng2015PRX,Hasan2015Science,Hasan2015NPhys,Ding2015NPhys,
WangFang2012PRB,Young2012PRL,Steinberg2014PRL,Hussain2014NMat,
LiuChen2014Science,Ong2015Science,Montambaux-Fuchs-PB2012} that own a number of
discrete Dirac points and a linear dispersion in two or three directions
irrespective of their microscopic details~\cite{Neto2009RMP,Hasan2010RMP,Qi2011RMP,
Huang2015PRX,Hasan2015Science,Hasan2015NPhys,Ding2015NPhys}, the two-dimensional (2D)
semi-Dirac (SD) electronic semimetals, one cousin of Dirac-like family, have recently been
attracting many studies~\cite{Hasegawa2006PRB,Pardo2009PRL,Katayama2006JPSJ,
Dietl2008PRL,Delplace2010PRB,Banerjee2009PRL,Montambaux-Piechon2009PRB,
Banerjee2012PRB,Montambaux-Fuchs2012PRL,Wu2014Expre,Saha2016PRB,
Uchoa2017PRB,Yang-Isobe2014NP,Isobe2016PRL,
Cho-Moon2016SR,WLZ2017PRB,Roy2018PRX,Wang2018JPCM,Quan2018JPCM}
attesting to the unique dispersion around their Dirac points, namely parabolic in one
direction and linear in the other. To be concrete, they were widely presented
in distinct circumstances, for instance, the quasi-two dimensional organic
conductor $\alpha-(\mathrm{BEDT-TTF)_2I_3}$ salt under uniaxial pressure~\cite{Katayama2006JPSJ},
tight-binding honeycomb lattices for the presence of a magnetic field~\cite{Dietl2008PRL},
and the $\mathrm{VO_2-TiO_2}$ multilayer systems (nanoheterostructures)~\cite{Pardo2009PRL}
as well as photonic systems consisting of a square array of elliptical dielectric cylinders~\cite{Wu2014Expre}. In principle, there are at least three major ingredients, which are
expected to be intimately associated with the low-energy fates of physical properties of fermionic
systems, for instance the ground states, transport quantities and so on~\cite{Sachdev1999Book,Altland2002PR,
Lee2006RMP,Neto2009RMP,Fradkin2010ARCMP,Hasan2010RMP,Sarma2011RMP,Qi2011RMP,
Kotov2012RMP}. Specifically, the first ingredient is the dispersion of low-energy
excitations and the second one is the kind of fermion-fermion interactions that glues
these low-energy excitations. Lastly, the potential impurity scattering serves
as the third ingredient, which is always present in real systems.

It is therefore of considerable significance to explore how these physical facets influence
the low-energy properties of 2D SD materials. One of the most interesting phenomena is the
development of superconductivity. The well-known Bardeen-Cooper-Schrieffer (BCS)
theory~\cite{BCS1957PR} tells us that an arbitrarily weak attractive force can glue
a pair of electrons and induce the Cooper pairing instability in normal metals, which
is directly linked to the superconductivity. This process can be
expressed alternatively by virtue of the language of modern
renormalization group theory, namely the absolute strength of
attractive interaction is (marginally) relevant with respect to the corresponding
effective model, which eventually runs to the strong (infinite) coupling no matter how
small its starting value is~\cite{Wilson1975RMP,Polchinski9210046,Shankar1994RMP}.
Recently, the Cooper pairing of Dirac fermions, in particular intrinsic Dirac semimetals,
has been paid a multitude of attentions~\cite{Neto2009RMP,Zhao2006PRL,Honerkamp2008PRL,
Roy-Herbut2010PRB,Roy-Herbut2013PRB,Roy-Jurici2014PRB,Ponte-Lee2014NJP,
Yao2015PRL,Maciejko2016PRL,Chubukov-Levitov2012NP,Sondhi2013PRB,Sondhi2014PRB}.
One of the most important points addressed  by previous works is that the Cooper pairing
only forms once the absolute value of attractive interaction exceeds certain
critical value owing to the vanishing density of states (DOS) and linear dispersions
at the Dirac points of Dirac semimetals (DSM). This implies the Cooper theorem does not
work and there may exist some quantum phase transition tuned by the strength of
attractive interaction~\cite{Zhao2006PRL,Honerkamp2008PRL,Sondhi2013PRB,Sondhi2014PRB}.

In comparison with the DSM, the 2D SD semimetals possess even more unconventional
features in that they harbor unusually anisotropic dispersions besides the zero DOS
at the discrete Dirac points~\cite{Banerjee2009PRL,Delplace2010PRB,Banerjee2012PRB}.
Motivated by all these considerations, it is consequently of remarkable interest to
explore whether the superconductivity accompanied by the Cooper instability can be
triggered once certain attractive fermion-fermion interaction is switched on in the 2D SD
materials and pin down the necessary requirements for this instability as well as
the influence caused by impurity scatterings, which are always inevitable and bring out two
converse contributions, namely both shortening lifetimes of quasi particles and enhancing
the DOS of fermions? Unambiguously elucidating these questions would be of remarkable
help for us to further fathom the unusual behaviors of 2D SD materials and
even profitable to seek new Dirac-like materials~\cite{LCJS2009PRL,Beenakker2009PRL,Rosenberg2010PRL,
Hasan2011Science,Bahramy2012NC,Viyuela2012PRB,Bardyn2012PRL,Garate2003PRL,
Oka2009PRB,Lindner2011NP,Gedik2013Science,CBHR2016PRL,Slager1802}.

In order to capture more physical information, we, on one hand, need to involve more
physical ingredients and on the other hand, take into account them unbiasedly in the
low-energy regime. To this end, a good candidate is the powerful renormalization group (RG)
approach~\cite{Wilson1975RMP,Polchinski9210046,Shankar1994RMP}.
To be specific, we within this work, besides the non-interacting Hamiltonian, will bring
out the Cooper-pairing interaction, which is obtained via performing the projection of
an attractive fermion-fermion interaction~\cite{Sondhi2013PRB,Sondhi2014PRB,Wang2017PRB_BCS}.
To proceed, we carefully investigate the effects
of this Cooper-pairing interaction and impurities as well as a nonzero
chemical potential on the emergence of Cooper instability in the
low-energy regime of 2D SD systems by virtue
of the RG approach.

In brief, our central focus is on whether and how the
Cooper instability can be generated. For completeness, we explicitly
study this problem at both zero and a finite chemical potential.  At first,
we consider the $\mu=0$ case. Conventionally, there are in all three types of
one-loop diagrams, namely ZS, $\mathrm{ZS}'$, and BCS~\cite{Shankar1994RMP},
contributing to the Cooper-pairing coupling $\lambda$~(\ref{Eq_H_int2}), whose
divergence is directly related to the Cooper instability.
In the 2D DSM systems, the BCS diagram is dominant and primarily responsible for the
Cooper instability (usually dubbed as the BCS instability due to
its leading contribution).
In a sharp contrast, the particular distinction from the 2D DSM materials is
that the BCS contribution vanishes for 2D SD systems at $\mu=0$. Unlike the BCS subchannel,
the RG running of parameter $\lambda$ can collect the corrections from both ZS and $\mathrm{ZS}'$ diagrams
once the internal transfer momentum $\mathbf{Q}$ is nonzero. After carrying out both analytical and
numerical analysis, we find that the Cooper theorem is invalid, i.e., Cooper instability cannot be
activated by any weak attractive fermioic interaction in 2D SD materials. However, once the starting
value of fermion-fermion coupling $\lambda$ goes beyond certain critical value,
it can be produced by the summation of ZS plus $\mathrm{ZS}'$, which is intimately
dependent upon the strength and direction of the transfer momentum $\mathbf{Q}$.
To be concrete, the Cooper instability cannot be ignited within some directions
of $\mathbf{Q}$ even its strength is large. However, it can be successfully
induced once the strength and direction of $\mathbf{Q}$ belong to a confined region
and the initial strength of $|\lambda(0)|$ exceeds the
certain critical strength. Next, we turn to the $\mu\neq0$ circumstance.
The one-loop RG analysis indicates that
the chemical potential $\mu$ is a relevant parameter, which is increased quickly via
lowering the energy scale. As a result, any weak Cooper-pairing interaction can
induce the Cooper instability, namely the Cooper theorem being restored~\cite{BCS1957PR}.
With this respect, one can expect a $\mu$-tuned phase transition associated with
the Cooper instability. Furthermore, the impurities play significant roles in determining the low-energy
properties of the real fermionic systems~\cite{Ramakrishnan1985RMP,Nersesyan1995NPB,Mirlin2008RMP,
Efremov2011PRB,Efremov2013NJP,Korshunov2014PRB,Fiete2016PRB,
Nandkishore2013PRB,Potirniche2014PRB,Nandkishore2017PRB,Stauber2005PRB,Roy2016PRB-2,
Roy1604.01390,Roy1610.08973,Roy2016SR,Aleiner2006PRL,Aleiner2006PRL-2,Lee1702.02742,Wang2011PRB}.
Concretely, they can both generate fermion excitations to suppress
the superconductivity and enhance the DOS of system
to be helpful for the superconductivity.  As the Cooper instability is
directly linked to the superconductivity, it is tempting to ask how the
impurity influences the stability of Cooper instability.
Concretely, we firstly study the influence of three primary
types of impurities on the Cooper instability at $\mu=0$, which
are named as random chemical potential, random mass, and random gauge
potential, respectively~\cite{Nersesyan1995NPB,Stauber2005PRB,Wang2011PRB,Wang2013PRB} and
distinguished by their distinct couplings with fermions presented in Eq.~(\ref{Eq_S_f-d}).
As the chemical potential and impurities scatterings contribute distinctly to the Cooper instability,
we, for completeness, also briefly examine whether and how the fate of the Cooper instability
is influenced by the competition between the impurities and a finite chemical potential.

We organize the rest parts of this work as follows. The Cooper-pairing interaction is introduced and
effective theory is constructed in Sec.~\ref{Sec_model}. We within Sec.~\ref{RG_clean_case}
compute the evaluations of one-loop diagrams and perform the standard RG analysis
to derive the coupled flow equations of interaction parameters. The Sec.~\ref{Sec_Discussion}
is accompanied to investigate whether and how the Cooper instability can be generated by the attractive
Cooper-pairing interaction at $\mu=0$ as well as the effects of a finite chemical potential.
In Sec.~\ref{Sec_imp}, we present a brief discussion on the stability of Cooper instability
against the impurity scatterings at $\mu=0$. The Sec.~\ref{Sec_imp-mu} is followed
to the effects of competition between impurities and a nonzero chemical potential.
Finally, a short summary is provided in Sec.~\ref{Sec_summary}.

\section{Effective theory}\label{Sec_model}

\subsection{Non-interacting model and Cooper-pairing interaction}

We employ the following non-interacting model
to capture the low-energy information of a 2D SD
system~\cite{Banerjee2009PRL,Montambaux-Piechon2009PRB,Delplace2010PRB,Saha2016PRB}
\begin{eqnarray}
\mathcal{H}_0(\mathbf{k})=(\alpha k^2_x-\delta)\sigma_1+vk_y\sigma_2,\label{Eq_H0}
\end{eqnarray}
with the parameters $\alpha$ and $v$ being respectively
the inverse of quasiparticle mass along $x$ and Dirac velocity along $y$, as well as
$\delta$ the gap parameter. Here $\sigma_1$ and $\sigma_2$ are Pauli matrixes.
Attesting to its unusual energy eigenvalues derived from Eq.~(\ref{Eq_H0}),
$E^{\pm}=\pm\sqrt{(\alpha k^2_x-\delta)^2+v^2k^2_y}$,
one can realize that the spectrum and ground state intimately rely upon the value of parameter
$\delta$. To be concrete, there exists
two gapless Dirac points at $(\pm\frac{\delta}{\alpha},0)$ while $\delta>0$ and
the system becomes a trivial insulator with a finite energy gap if
$\delta<0$. In a sharp contrast, the spectrum is gapless with the linear
dispersion along $k_y$ and parabolical for $k_x$ directions
at $\delta=0$.

Without loss of generality, we within this work focus on the first case ($\delta=0$)
due to the peculiarly anisotropic dispersion along
$k_x$ and $k_y$ orientations. Additionally, the effects of chemical potential on the low-energy states
would be examined. Gathering these considerations together, we expand the dispersion in the vicinity
of the Dirac point and accordingly arrive at the non-interacting effective
action~\cite{Saha2016PRB,Uchoa2017PRB,Wang2018JPCM}
\begin{eqnarray}
S_0&=&\int\frac{d\omega}{(2\pi)}\frac{d^2\mathbf{k}}{(2\pi)^2}\Psi^\dagger(i\omega,\mathbf{k})
(-i\omega+\alpha k^2_x\sigma_1\nonumber\\
&&+vk_y\sigma_2-\mu)\Psi(i\omega,\mathbf{k}).\label{Eq_S_0}
\end{eqnarray}
Here, the $\sigma_i$, with $i=1,2,3$ again corresponds to the Pauli matrices, which satisfy
the algebra $\{\sigma_i,\sigma_j\}=2\delta_{ij}$. In addition, the spinors $\Psi^\dagger(i\omega,\mathbf{k})$
and $\Psi(i\omega,\mathbf{k})$ specify the low-energy excitations of fermionic degrees from the Dirac point.
In accordance with this non-interacting model~(\ref{Eq_S_0}), the free fermionic propagator can be
straightforwardly extracted as
\begin{eqnarray}
G_0(i\omega,\mathbf{k})
&=&\frac{1}{-i\omega+\alpha k^2_x\sigma_1+vk_y\sigma_2-\mu}.\label{Eq_G_0}
\end{eqnarray}
Further, we stress that the parameter $\mu$ refers to the chemical potential whose effects on
the low-energy physics will be studied in next sections.

\begin{figure}
\centering
\includegraphics[width=3.0in]{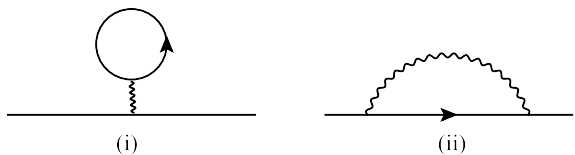}
\vspace{-0.1cm}
\caption{One-loop corrections to the fermionic propagator at clean limit due to the
Cooper-pairing interaction. The wave lines denote the Cooper-pairing
interaction. Notice that one-loop corrections from the fermion-impurity
interaction can be obtained via replacing the wave lines with
dashed lines describing the fermion-impurity interaction explicitly shown
in Fig~\ref{Fig2_fermion_propagator_correction_imp}.}\label{Fig1_fermion_propagator_correction}
\end{figure}

We would like to point out one of the main purposes within this work is to explore the distinct
behaviors of low-energy states in the 2D SD semimetals between zero and finite chemical
potential as the density of states at Dirac point is qualitatively changed.
In this respect, one can directly let $\mu=0$ and utilize
the corresponding propagator while it is necessary.

\subsection{Cooper-pairing interaction}

Besides the non-interacting action, we subsequently bring out an attractive fermion-fermion interaction
~\cite{Sondhi2013PRB,Sondhi2014PRB,Wang2017PRB_BCS},
\begin{eqnarray}
\mathcal{H}_{\mathrm{int}}
&=&\int d^2{\mathbf{r}}\frac{\lambda(\mathbf{r})}{4}\Psi^\dagger(\mathbf{r})\Psi(\mathbf{r})
\Psi^\dagger(\mathbf{r})\Psi(\mathbf{r}),\label{Eq_H_int}
\end{eqnarray}
with the coupling strength function $\lambda(\mathbf{r})<0$. To simplify our analyses, we assume
$\lambda(\mathbf{r})$ to be a constant initially and run upon lowering the energy scale after
taking into account the higher-order corrections.

\begin{figure}
\centering
\includegraphics[width=0.9in]{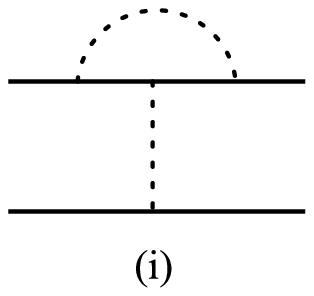}\hspace{0.368cm}
\includegraphics[width=0.9in]{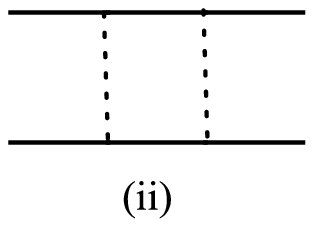}\hspace{0.368cm}
\includegraphics[width=0.9in]{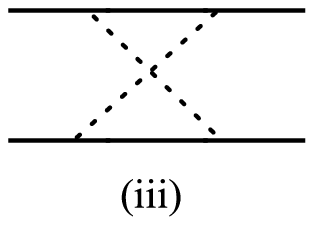}\\ \vspace{0.68cm}
\includegraphics[width=0.9in]{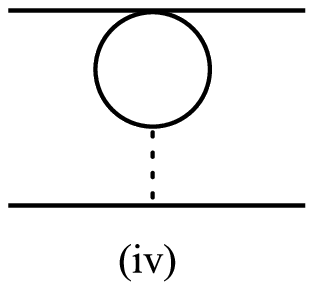}\hspace{0.3cm}
\includegraphics[width=0.9in]{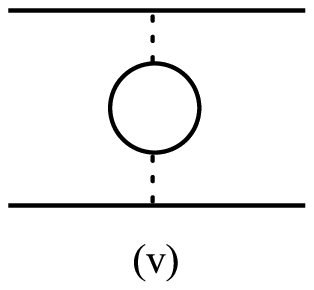}
\vspace{-0.1cm}
\caption{One-loop corrections to the fermion-impurity strength. 
Subfigures (i)-(v) denote distinct sorts of topological structures. The dashed lines specify
fermion-impurity interaction.}\label{Fig2_fermion_propagator_correction_imp}
\end{figure}

To proceed, we are going to start manifestly from an effective Cooper-pairing interaction (only
focusing on the singlet pairing here), which involves only the pairing between two fermions that
carry both opposite momenta and spin directions. In order to realize this, we, referring to the
approach by Nandkishore et \emph{al.}~\cite{Sondhi2013PRB}, try to perform the projection of the full
interaction~(\ref{Eq_H_int}) onto the Cooper-pairing channel. To be specific, one needs to firstly
translate the interaction~(\ref{Eq_H_int}) into its momentum-space
version via performing a Fourier transformation and next bring out
a delta function $\delta^2(\mathbf{k}_1+\mathbf{k}_2)$ to the updated
interaction and finally integrate the momenta $\mathbf{k}_2$ out.
After fulfilling these procedures, the Cooper-pairing interaction can be formally achieved, namely
\begin{eqnarray}
\mathcal{H}_{\mathrm{Coop}}
\!=\!\!\!\sum_{\mathbf{k}_1,\mathbf{k}_2}\!\!\frac{\lambda\Lambda^2}{4}
\Psi^\dagger_{\mathbf{k}_1,\uparrow}(-i\sigma_2)\Psi^\dagger_{-\mathbf{k}_1,\downarrow}
\Psi_{-\mathbf{k}_2,\downarrow}(i\sigma_2)\Psi_{\mathbf{k}_2,\uparrow},\label{Eq_H_int2}
\end{eqnarray}
which will be regarded as our starting point of effective interaction. However, one central
point we have to highlight is that the delta function $\delta^2(\mathbf{k})$ scales like
$\mathbf{k}^{-2}$, which is added by hand during the process for
deriving the Cooper interaction. Consequently, the dimension of fermionic
coupling $\lambda$ would be changed. To remedy this, we bring about an UV cutoff $\Lambda$
to above effective interaction, which can be understood as a scaling to provide the
corresponding dimensions. Without loss of generality, we will make the transformation $\lambda\Lambda^2/4\rightarrow\lambda$ in our analyses of next sections.

\subsection{Fermion-impurity interaction and effective theory}\label{Subsec_imp}

We hereby only focus the study on a quenched, Gauss-white potential under the conditions~\cite{Nersesyan1995NPB,Stauber2005PRB,Wang2011PRB,Mirlin2008RMP,Altland2006Book,Coleman2015Book},
whose impurity field $\mathcal{I}$ satisfies the following restrictions
\begin{eqnarray}
\langle \mathcal{I}(\mathbf{x})\rangle=0,\hspace{0.5cm}\langle \mathcal{I}(\mathbf{x})\mathcal{I}(\mathbf{x'})\rangle
=\Delta\delta^2(\mathbf{x}-\mathbf{x'}),\label{Eq_S_d-d}
\end{eqnarray}
where the parameter $\Delta$ specifies the concentration of the impurity and can be taken as a constant
controlled by the experiments.

We bring out the fermion-impurity interaction~(scattering) via adopting the
replica technique~\cite{Edwards1975JPF,Ramakrishnan1985RMP,
Mirlin2008RMP,Wang2015PLA} to average over the random impurity potential $\mathcal{I}(\mathbf{x})$,
\begin{eqnarray}
S_{\mathcal{I}}\!\!&=&\!\!\sum_\mathrm{I}\frac{\Delta_I}{2}\int\prod^{l=2,l'=3}_{l=1,l'=1}
\frac{d\omega_l d^2\mathbf{k}_l'}{(2\pi)^8}\Psi^\dagger_m(\omega_1,\mathbf{k}_1)
\gamma_\mathrm{I}\Psi_m(\omega_1,\mathbf{k}_2)\nonumber\\
&&\times\Psi^\dagger_n(\omega_2,\mathbf{k}_3)
\gamma_\mathrm{I}\Psi_n(\omega_2,\mathbf{k}_1+\mathbf{k}_2-\mathbf{k}_3),\label{Eq_S_f-d}
\end{eqnarray}
where the parameters $m$ and $n$ describe the two replica indexes and the parameter
$\Delta_I=\Delta v^2_\mathrm{I}$ with $\mathrm{I}$
being $C$, $M$, $G_{1,3}$ to distinguish different sorts of impurities one by one,
which will be utilized to specify the strength of impurity scattering and the coupling
$v_\mathrm{I}$ characterizing the strength of a single impurity~\cite{Nersesyan1995NPB,Coleman2015Book}.
The Pauli matrix $\gamma_\mathrm{I}$ respectively corresponds three typical sorts of impurities, which
are dubbed by random chemical potential~($\gamma=\sigma_0$), random mass~($\gamma=\sigma_2$),
and random gauge potential~($\gamma=\sigma_{1,3}$)~\cite{Nersesyan1995NPB,Stauber2005PRB}.

\begin{figure}
\centering
\includegraphics[width=3.3in]{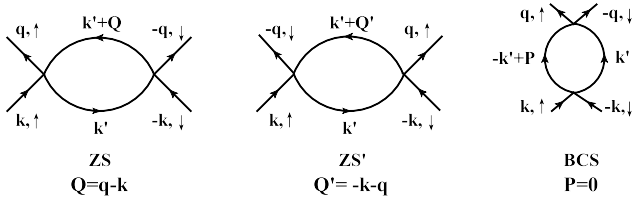}
\vspace{-0.0cm}
\caption{One-loop corrections to the attractive Cooper-pairing coupling from ZS,
$\mathrm{ZS}'$, and BCS subchannels.}\label{Fig3_ZS_BCS}
\end{figure}

Gathering non-interacting Hamiltonian and attractive Cooper-pairing interaction as well as
fermion-impurity interaction together, we
subsequently arrive at the effective theory that contains the Cooper
channel and the fermion-impurity interactions,
\begin{widetext}
\begin{eqnarray}
S_{\mathrm{eff}}&=&\int\frac{d\omega}{2\pi}\int\frac{d^{2}\mathbf{k}}{(2\pi)^2}
\Psi^\dagger(i\omega,\mathbf{k})
\left(-i\omega+\alpha k^2_x\sigma_1+vk_y\sigma_2-\mu\right)\Psi(i\omega,\mathbf{k})
+\left(\frac{\lambda\Lambda^2}{4}\right)\int\frac{d\omega_1d\omega_2d\omega_3}
{(2\pi)^3}\int\frac{d^{2}\mathbf{k}_1d^{2}\mathbf{k}_2}{(2\pi)^4}\nonumber\\
&&\times\Psi^\dagger(i\omega_1,\mathbf{k}_1,\uparrow)(-i\sigma_2)
\Psi^\dagger(i\omega_2,-\mathbf{k}_1,\downarrow)
\Psi(i\omega_3,-\mathbf{k}_2,\downarrow)(i\sigma_2)
\Psi(i\omega_1+i\omega_2-i\omega_3,\mathbf{k}_2,\uparrow)
+S_{\mathcal{I}}.\label{Eq_S_eff}
\end{eqnarray}
\end{widetext}
To be consistent, we at the moment address short comments on the possibility of
this attractive Cooper channel interaction~(\ref{Eq_H_int2}).
Generally, electron-electron interaction is repulsive owing to the Coulomb
interaction. Fortunately, the attractive interactions can be switched on via
either phonons or plasmons~\cite{Uchoa2007PRL,Neto2009RMP}.
As a result, an essential problem is to reduce or screen the Coulomb interaction.
Despite the full screening of repulsive Coulomb interaction in 2D SD system is hardly
realized at $\mu=0$~\cite{Katsnelson2006PRB}, one can partially or even substantially
suppress the Coulomb interaction via adopting some metallic substrate
to increase the dielectric constant~\cite{Neto2009RMP,Kotov2012RMP,Sarma2011RMP}.
Further, the chemical potential that qualitatively
changes the Dirac point and generates a finite DOS at fermi surface may also greatly
suppresses the Coulomb interaction. Therefore, a net attractive interaction is allowed
once the absolute strength of Coulomb interaction is smaller than its attractive counterpart.
With these respects, it is in principle possible to form a net attractive interaction for our system.
Our impeding study will be based on the assumption that a net attractive force is realized.

\begin{figure}
\centering
\includegraphics[width=1.5in]{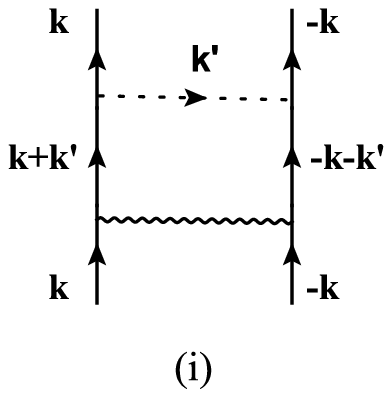}\hspace{0.6cm}
\includegraphics[width=1.5in]{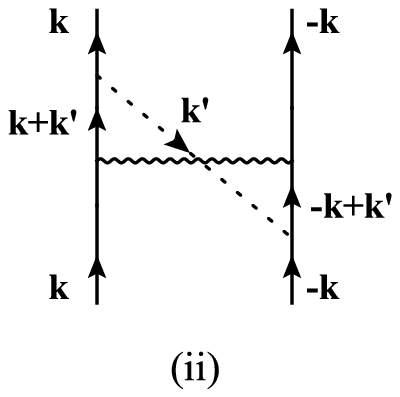}
\vspace{-0.0cm}
\caption{One-loop corrections to the Cooper-pairing coupling $\lambda$ due to
the fermion-impurity interactions. Subfigures (i) and (ii) correspond to particle-particle and
particle-hole channels, respectively.}\label{Fig4_lambda_correction_imp}
\end{figure}

Reading off our effective theory~(\ref{Eq_S_eff}), it is of remarkable interest
to stress that the attractive Cooper channel interaction~(\ref{Eq_H_int2}) can
generate three sorts of one-loop diagrams, namely, ZS, $\mathrm{ZS}'$ and BCS,
which all contribute to the coupling strength $\lambda$ and together play an
important role in determining low-energy behaviors~\cite{Shankar1994RMP,Sondhi2013PRB,Sondhi2014PRB}. Accordingly, the low-energy properties of 2D SD semimetals, in particular whether the
Cooper instability can be ignited, are primarily governed by these one-loop corrections from
fermionic attractive interaction together with the chemical potential $\mu$. In order to examine
this within a wide energy regime, we are suggested to derive energy-dependent evolutions of
interaction parameters and investigate the low-energy behaviors by virtue of unbiased
renormalization group approach, which can treat all potential facets on the same
footing and thus capture the mutual effects among all interaction parameters.
In this work, we concentrate on one-loop corrections, which are related to Feynman diagrams provided in Figs.~\ref{Fig1_fermion_propagator_correction}-\ref{Fig4_lambda_correction_imp}
respectively stemming from Cooper-pairing~(Figs.~\ref{Fig1_fermion_propagator_correction},
\ref{Fig3_ZS_BCS}) and fermion-impurity interaction~(Figs.~\ref{Fig1_fermion_propagator_correction},
\ref{Fig2_fermion_propagator_correction_imp}, \ref{Fig4_lambda_correction_imp}).

\section{Renormalization-group analysis at clean limit}\label{RG_clean_case}

In this section, we only concentrate on the clean-limit case, namely neglecting
$S_{\mathcal{I}}$ in Eq.~(\ref{Eq_S_eff}), and leave the analysis in presence of
impurity in Sec.~\ref{Sec_imp}. To be specific, we complete the one-loop RG analysis
of effective theory~(\ref{Eq_S_eff}) to construct the coupled running equations of all
correlated parameters upon lowering the energy scales via adopting the momentum-shell RG method~\cite{Wilson1975RMP,Polchinski9210046,Shankar1994RMP}. Along with the standard steps
of this RG framework, one integrates out the fast modes of fermionic
fields characterized by the momentum shell $b\Lambda<k<\Lambda$ with the variable parameter
$b=e^{-l}<1$ and a running energy scale $l$, then incorporates these fast-mode contributions
to the slow modes, and finally rescales the slow modes to new ``fast modes"~\cite{Huh2008PRB,Kim2008PRB,
Chubukov2010PRB,She2010PRB,She2015PRB,Vafek2012PRB,Vafek2014PRB,
Roy2016PRB,Wang2011PRB,Wang2013PRB,Wang2017PRB_QBCP}.
After performing all these procedures, the coupled flow RG equations of
interaction parameters can be derived via comparing new ``fast modes"
with old ``fast modes" in the effective theory.

These coupled flow equations of all interaction parameters are generally pivotal
to determine the low-energy physical behaviors. Before moving further, we need to
derive the RG rescaling transformations of fields and momenta,
which connect two continuous steps of RG processes. In accordance with
the spirit of the momentum-shell RG, the non-interacting parts
$(-i\omega+\alpha k^2_x\sigma_1+vk_y\sigma_2)$ can be conventionally selected as a
starting fixed point, which is invariant during the RG transformations. Under these
respects, the RG re-scaling transformations can be extracted as~\cite{Shankar1994RMP,Huh2008PRB,Wang2011PRB},
\begin{eqnarray}
k_{x}&=&k'_{x}e^{-\frac{1}{2}l},\label{Eq_rescaling_kx}\\
k_{y}&=&k'_{y}e^{-l},\\
\omega&=&\omega'e^{-l},\\
\Psi(i\omega,\mathbf{k})
&=&\Psi'(i\omega',\mathbf{k}')e^{\frac{1}{2}\int^l_0dl\left(\frac{7}{2}-\eta\right)},\label{Eq_rescaling_psi}
\end{eqnarray}
where the parameter $\eta$ is closely linked to the higher-level corrections due to
the fermioinc interactions, which characterizes the potentially anomalous dimension of fermionc spinor.
It is worth pointing out that these re-scalings can be understood as the bridge between
the ``old" and ``new" fast modes of the effective theory, which would play a vital role
in building the coupled RG evolutions of all related interaction parameters.

At this stage, we consequently can concentrate on our RG analyses. As delineated in
Eq.~(\ref{Eq_S_eff}), there are in all four parameters that we need to care about,
namely $\alpha$, $v$, $\mu$, and $\lambda$. To proceed, we begin with the tree-level case
at which we turn off the higher-order corrections. After considering the re-scalings from Eq.~(\ref{Eq_rescaling_kx})-Eq.~(\ref{Eq_rescaling_psi}),
one can straightforwardly find the evolutions as follows
\begin{eqnarray}
\frac{d\mu}{dl}&=&\mu,\label{Eq_RG_mu_0}\\ \nonumber \\
\frac{d\lambda}{dl}&=&-\lambda,\label{Eq_RG-lambda-tree}
\end{eqnarray}
and the parameters $d\alpha/dl=dv/dl=0$. Under this situation, the interaction parameters
are evolving independently with decreasing the energy scale. As a result, the correlated
low-energy physical behaviors of 2D SD systems cannot be extracted and displayed. In particular, the Cooper
instability is directly forbidden by the RG equation of coupling $\lambda$~(\ref{Eq_RG-lambda-tree}).

In order to capture more physical information and pin down the fate of attractive interaction
$\lambda$ in the low-energy regime, we are forced to study the one-loop corrections to
the fermionic propagator and strength of fermionic interaction owing
to the attractive fermionic interaction. Before going further, we measure the momenta and energy with the cutoff $\Lambda_0$, which is associated with the lattice constant, namely $k\rightarrow k/\Lambda_0$ and $\omega\rightarrow\omega/\Lambda_0$~\cite{Shankar1994RMP,Huh2008PRB,She2010PRB,
Wang2011PRB}. According to one-loop corrections as depicted in Fig.~\ref{Fig1_fermion_propagator_correction} to fermionic propagator, there exists no anomalous fermionic dimension, namely, $\eta=0$.

In addition, we turn to the one-loop corrections to interaction parameter $\lambda$,
which contain three distinct types of subchannels, namely ZS, $\mathrm{ZS}'$, and BCS subchannels~\cite{Shankar1994RMP} as delineated in Fig.~\ref{Fig3_ZS_BCS}.
Although both ZS and $\mathrm{ZS}'$ diagrams own a finite transfer momentum $\mathbf{Q}=\mathbf{q}-\mathbf{k}$
and $\mathbf{Q}'=-\mathbf{q}-\mathbf{k}$, it is of particular interest for
Cooper interaction to point out that
$|\mathbf{Q}|\ll|\mathbf{Q}'|$ once two external momenta $\mathbf{q}$ and $\mathbf{k}$ possess the same sign
(or $|\mathbf{Q}'|\ll|\mathbf{Q}|$ if they own opposite signs).
For simplicity, we can approximately let $\mathbf{Q}=0$ and take a finite value of $\mathbf{Q}'$ or vice
versa~\cite{Shankar1994RMP,Wang2017PRB_BCS}. Within this work, we also adopt this approximation
to simplify our analyses. To be specific, we assume $\mathbf{Q}=0$ and $\mathbf{Q}'$
acquires a finite value, which is characterized by two parameters $Q$ and $\varphi$
to measure its strength and direction, respectively. Carrying out several tedious
but straightforward calculations gives rise to the following
corrections,
\begin{eqnarray}
\delta\lambda_{\mathrm{ZS}}\!&=&\!\frac{\lambda^2l(8\mathcal{D}_1
-4\mu^2\mathcal{D}_0)}{4\pi^2},\label{Eq_one-loop-ZS}\\
\delta\lambda_{\mathrm{ZS'}}\!&=&\!\frac{\lambda^2l
\left[8(\mathcal{D}_2-\mathcal{D}_1-\sum^5_{i=3}\mathcal{D}_i)
+4\mu^2\mathcal{D}_0\right]}{4\pi^2},\label{Eq_one-loop-ZS-prime}\\
\delta\lambda_{\mathrm{BCS}}\!&=&\!\frac{2\lambda^2l\mu^2\mathcal{D}_0}{4\pi^2},\label{Eq_one-loop-BCS}
\end{eqnarray}
where the related coefficients $\mathcal{D}_i$
with $i=0$ to $5$ are designated in Eqs.~(\ref{Eq_D_1})-(\ref{Eq_D_5}). 
We hereby emphasize the one-loop corrections at $\mu=0$ can be calculated analogously, which
will be studied in details in Sec.~\ref{Sec_mu0}. Based on these one-loop corrections, we
derive the coupled RG evolutions at $\mu\neq0$:
\begin{eqnarray}
\frac{d\mu}{dl}\!\!&=&\!\!\!\mu,\label{Eq_RG_mu}\\
\frac{d\lambda}{dl}\!\!&=&\!\!\!\left[-1-\frac{\lambda\left(4\mathcal{D}_2-4\sum^5_{i=3}\mathcal{D}_i
+\mu^2\mathcal{D}_0\right)}{4\pi^2}\right]\!\!\lambda,\label{Eq_RG_lambda}
\end{eqnarray}
where the interaction parameters $d\alpha/dl=dv/dl=0$.

Before moving further, we now would like to present brief remarks
on these coupled RG evolutions of interaction parameters. At first,
one-loop RG evolution~(\ref{Eq_RG_lambda}) is qualitatively distinct
from their tree-level counterpart~(\ref{Eq_RG-lambda-tree}),
namely an additional term is generated no matter $\mu=0$ or $\mu\neq0$, which
may totally change its low-energy behaviors. This implies that
these couplings are not independent and hence their low-energy fates are intimately
associated with each other. Accordingly, the low-energy behaviors, compared to
their tree-level situations, may be revised or even qualitatively changed.
In particular, the fate of parameter $\lambda$ may be changed and Cooper instability
may be triggered under certain circumstance. In addition, the coupled RG
running equations are of remarkable distinction between zero and a finite chemical potential
as the values of DOS at the Dirac point are qualitatively different. One therefore
can expect the distinct fates of the interaction coupling $\lambda$ between these
two cases, which may correspond to some phase transition.  Moreover, what about
the behaviors of physical quantities while the system undergoes a potentially tuned
phase transition? Whether the Cooper instability can be generated?
In the impending sections, we are going to study and response to these questions.

\section{Cooper instability and $\mu$-tuned phase transition at clean limit}\label{Sec_Discussion}

Within this section, we endeavor to investigate the effects of attractive Cooper-pairing
interaction and chemical potential on the low-energy behaviors of interaction
coupling $\lambda$ by virtue of both theoretical and numerical analyses of the one-loop
RG evolutions, which are established in Sec.~\ref{RG_clean_case} and 
intertwine all related interaction parameters together
and . Based on these information, we would
examine whether the Cooper instability can be triggered at $\mu=0$ and what
conditions are required to trigger the Cooper instability for our 2D SD systems.
In addition, since the DOS at Fermi surface (Dirac point) is qualitatively distinct between
$\mu=0$ and $\mu\neq0$ in 2D SD semimetals~\cite{Neto2009RMP,Banerjee2009PRL,Banerjee2012PRB},
one may expect a chemical potential-tuned ($\mu$-tuned) phase transition, which is
conventionally accompanied by unique behaviors in the vicinity of the critical point, for instance,
the Cooper instability attesting to its sensitivity to the Dirac point.

\subsection{Cooper instability at $\mu=0$}\label{Sec_mu0}

At the outset, we recall the tree-level results on the interaction coupling $\lambda$ depicted
in Eq.~(\ref{Eq_RG-lambda-tree}). In particular, we highlight that it flows independently with
parameters $\alpha$ and $v$ upon decreasing the energy scales. Accordingly, once
the parameter $\lambda$ is taken an initially attractive value, we can easily
find that Cooper instability cannot be activated as $\lambda$ goes towards zero
upon lowering energy scale. This indicates that the Cooper theorem is manifestly violated.
One may mainly ascribe this unusual feature to the
vanish of density of states at the Dirac point~\cite{Banerjee2009PRL,Delplace2010PRB,Banerjee2012PRB}.

In the spirit of RG theory, the higher-order corrections
are required to be involved to judge the stability of tree-level conclusion and
further pin down the fate of $\lambda$ at the low-energy region. To this end,
we calculate the one-loop contributions to the parameter $\lambda$, which are
consist of three subtypes, i.e., ZS, $\mathrm{ZS}'$ and BCS channels~\cite{Shankar1994RMP} as listed in
Eqs.~(\ref{Eq_one-loop-ZS})-(\ref{Eq_one-loop-BCS}).
To be concrete, These one-loop
corrections at $\mu=0$ are
derived as
\begin{eqnarray}
\delta\lambda_{\mathrm{ZS}}&=&\frac{8\lambda^2l\mathcal{D}_1}{4\pi^2},\label{Eq_one-loop-ZS-2}\\
\delta\lambda_{\mathrm{ZS'}}&=&\frac{8\lambda^2l
\left(\mathcal{D}_2-\mathcal{D}_1-\sum^5_{i=3}\mathcal{D}_i\right)}{4\pi^2},\label{Eq_one-loop-ZS-prime-2}\\
\delta\lambda_{\mathrm{BCS}}&=&0.\label{Eq_one-loop-BCS-2}
\end{eqnarray}
One needs to bear in mind during the derivations
that there are qualitative distinctions between 2D DSM, which possess linear dispersions
for both $k_x$ and $k_y$ directions, and our 2D SD semimetals. Accordingly,
the coupled evolutions of interaction parameters are derived as follows,
\begin{eqnarray}
\frac{d\alpha}{dl}&=&\frac{dv}{dl}=0,\label{Eq_RG_alpha-2}\\
\frac{d\lambda}{dl}&=&\left[-1-\frac{\lambda\left(4\mathcal{D}_2-4\sum^5_{i=3}\mathcal{D}_i\right)}
{4\pi^2}\right]\lambda.\label{Eq_RG_lambda-2}
\end{eqnarray}
Before moving further, we again stress that both one-loop corrections~(\ref{Eq_one-loop-ZS-2})-(\ref{Eq_one-loop-BCS-2})
and RG equations~(\ref{Eq_RG_alpha-2})-(\ref{Eq_RG_lambda-2}) are calculated and
derived separately.

\begin{figure}
\centering
\includegraphics[width=4.72in]{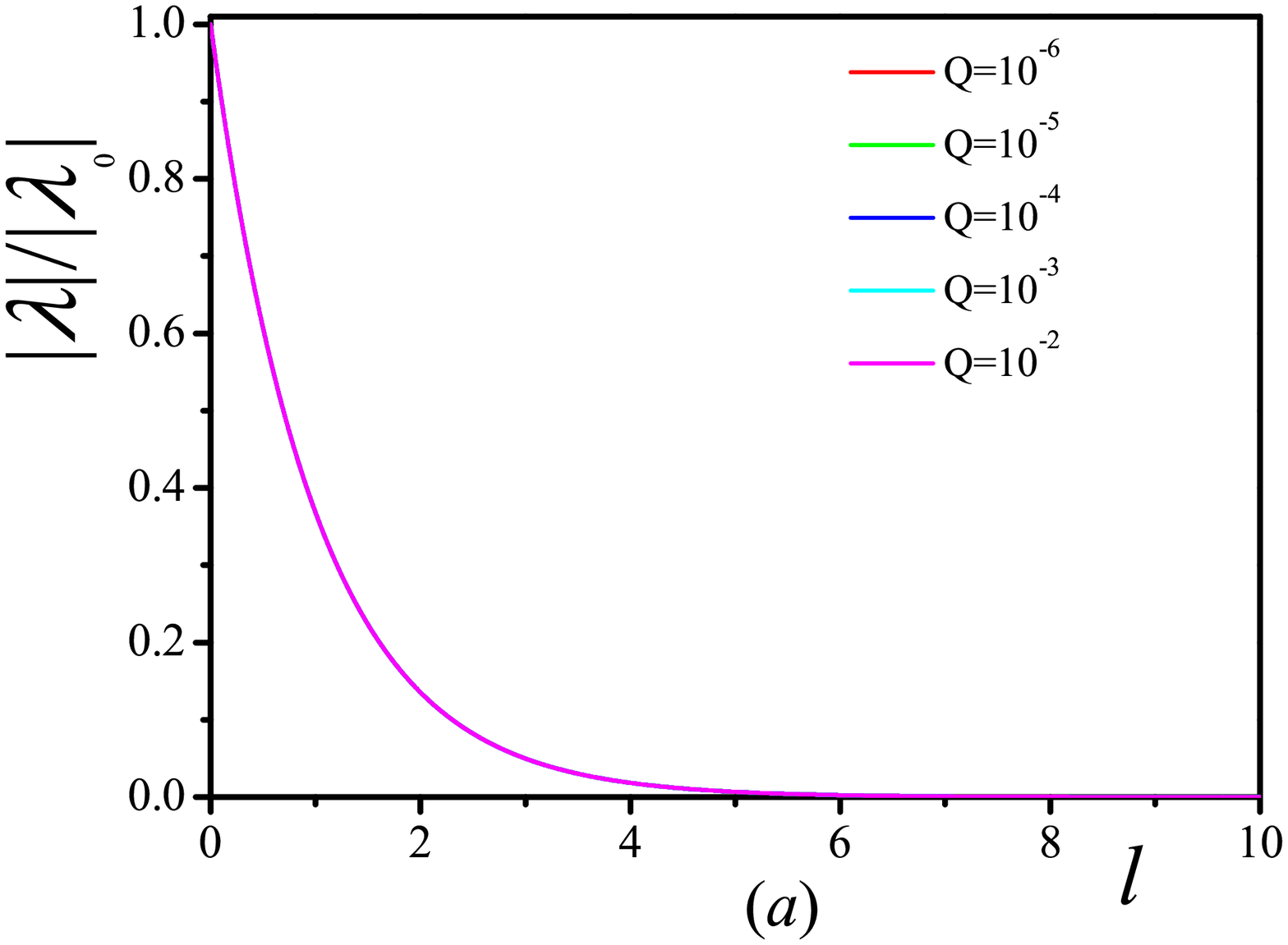} \\ \vspace{-2.05cm}
\includegraphics[width=4.72in]{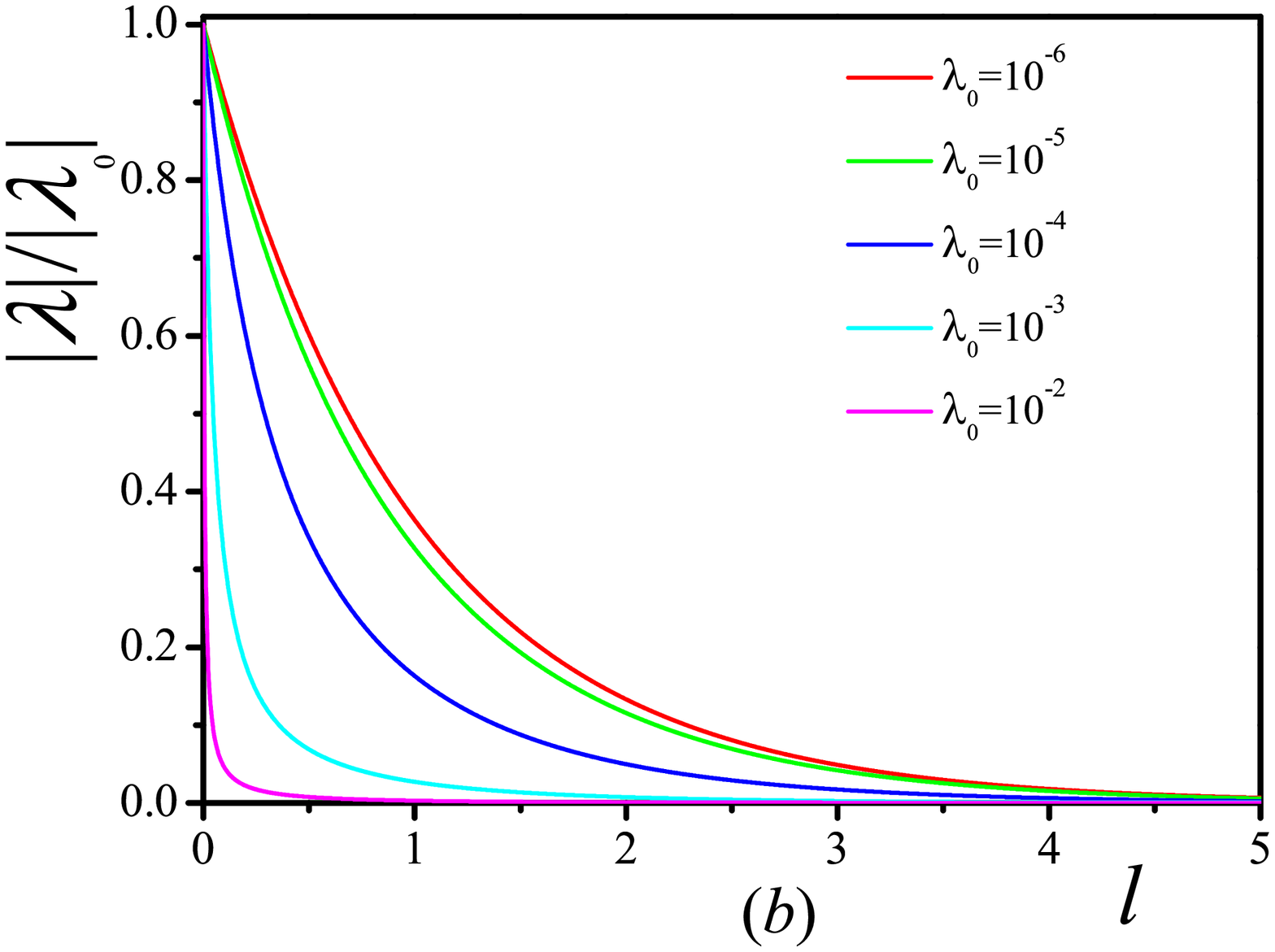}
\vspace{-2.1cm}
\caption{(Color online) Evolutions of $|\lambda|$ upon lowering energy scales
for $\mu=0$ and $\alpha(0)=5\times10^{-3}$,  $v(0)=10^{-3}$: (a). $\lambda(0)=-10^{-4}$ and $\varphi=\pi/2$
with several representative values of $Q$ and (b). $Q=10^{-3}$ and
$\varphi=\pi/3$ with several representative values of $\lambda_0\equiv\lambda(0)$.
Note the values of $\lambda$ and $Q$ are adequate to produce
the Cooper instability at other angels as shown in Fig.~\ref{Fig_mu0_lambda-Q_neq}.}
\label{Fig_mu0_lambda-Q-pi_neq}
\end{figure}

Learning from Eqs.~(\ref{Eq_one-loop-ZS-2})-(\ref{Eq_one-loop-BCS-2}),
it is of particular interest to point out that the BCS subchannel of Cooper-pairing interaction
does not contribute any corrections to the interaction coupling $\lambda$.
As a consequence, this subchannel does not participate in the coupled RG evolutions and
contribute to potential emergence of Cooper instability. This exhibits a sharp contrast
to the situation of 2D DSM materials, at which the BCS subchannel plays a central
role in igniting the Cooper instability (also dubbed as the BCS instability
owing to its leading contribution)
if the initial value of Cooper coupling exceeds certain critical value~\cite{Sondhi2013PRB,Wang2017PRB_BCS}.
We would like to pause hereby and remark on the underline logic that is responsible for their differences.
In brief, the cardinal facet is ascribed to the distinct dispersions of
low-energy fermionic excitations. In the BCS subchannel, the transfer momentum is
zero, namely $\mathbf{Q}=0$ and thus its correction is proportional to
$\mathrm{Tr}(\sigma_2G(i\omega,\mathbf{k})\sigma_2G(i\omega,-\mathbf{k}))$.
With respect to the 2D DSM systems, their dispersions are linear for both $k_x$ and $k_y$
directions, i.e., $G^{-1}(i\omega,\mathbf{k})\sim (-i\omega+c_1k_x\sigma_1+c_2k_y\sigma_2)$ with
$c_{1}$ and $c_2$ being some constants. As a result, corrections from $k_x$ and $k_y$ parts are mutually
neutralized each other and thus the $\omega$ term gains a finite contribution. Compared manifestly
to the 2D DSM's dispersion, our 2D SD semimetals possess anisotropic excitations along $k_x$
and $k_y$ orientations, namely $G^{-1}_{\mathrm{SD}}(i\omega,\mathbf{k})\sim (-i\omega+c_1k^2_x\sigma_1+c_2k_y\sigma_2)$. This consequently renders that $k_x$ and $k_y$
corrections support each other and finally their summation counteracts with the
correction from $(-i\omega)$ part. It therefore leads to the vanish of BCS subchannel at $\mu=0$.

We next turn to the contributions from the ZS and $\mathrm{ZS}'$ subchannels.
Specifically, we find that both ZS and $\mathrm{ZS}'$ diagrams can
contribute to the RG running of parameter $\lambda$ once the transfer momentum $\mathbf{Q}$
is nonzero. An exception is that the summation of ZS and $\mathrm{ZS}'$ subchannels can be neutralized
exactly in the case of $\mathbf{Q}=0$. According to the information above, we obtain that the coupling $\lambda$'s flow equation~(\ref{Eq_RG_lambda-2}) at $\mu=0$ only collects
the contributions from ZS and $\mathrm{ZS}'$ diagrams. This indicates that, at zero chemical potential,
the energy-dependent evolution of coupling $\lambda$ primarily hinges upon
the ZS plus $\mathrm{ZS}'$
not BCS subchannels, to be more specifically, the transfer momenta $\mathbf{Q}$.
As a consequence, it is tempting to ask whether the one-loop corrections from ZS and $\mathrm{ZS}'$
diagrams due to the Cooper-pairing interaction can produce the Cooper instability and how
it is related to the transfer momentum $\mathbf{Q}$.

\begin{figure}
\centering
\includegraphics[width=4.6in]{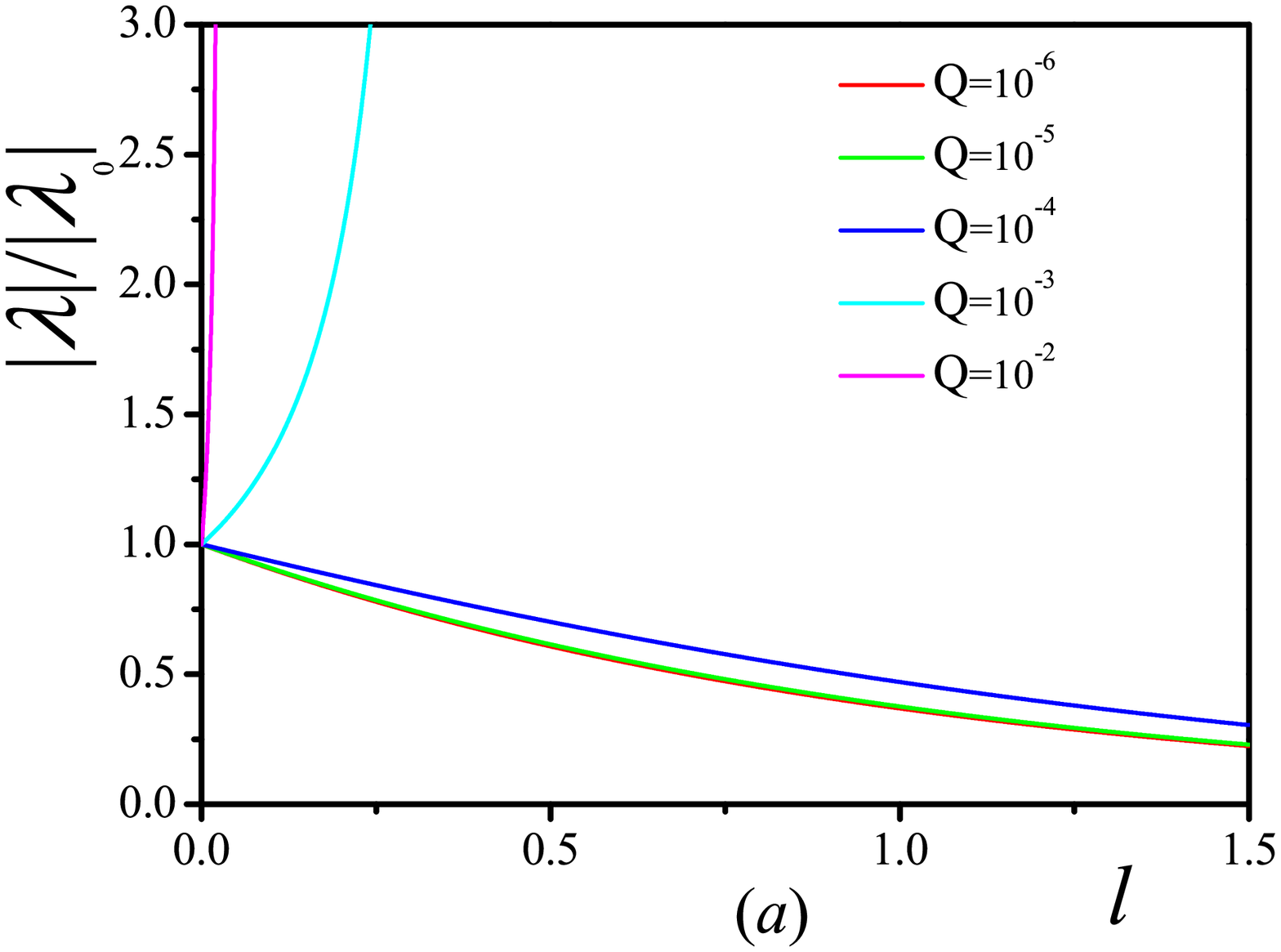}
\\ \vspace{-6.7cm}
\hspace{-0.1cm}\includegraphics[width=1.8in]{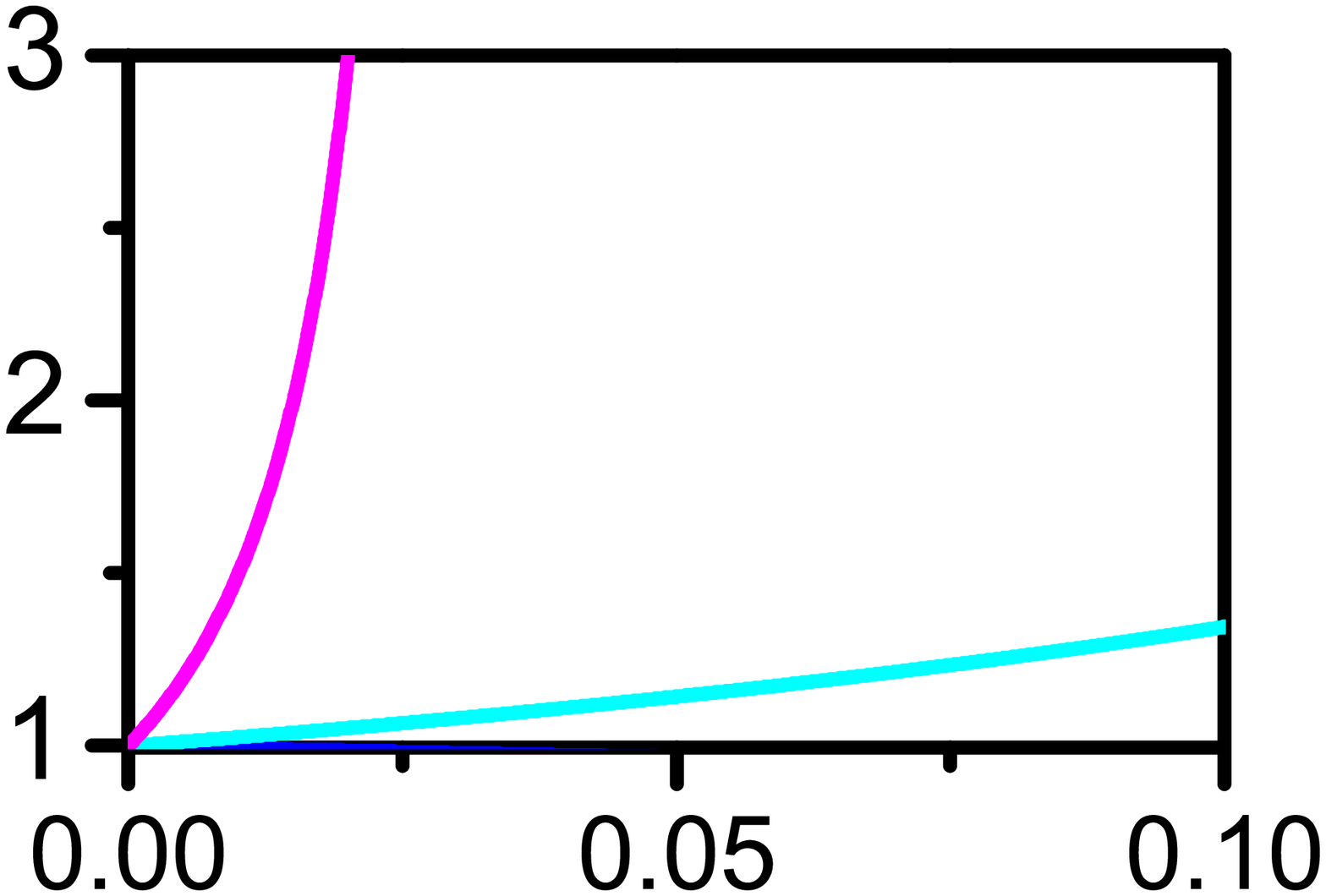}
\\ \vspace{2.0cm}
\includegraphics[width=4.6in]{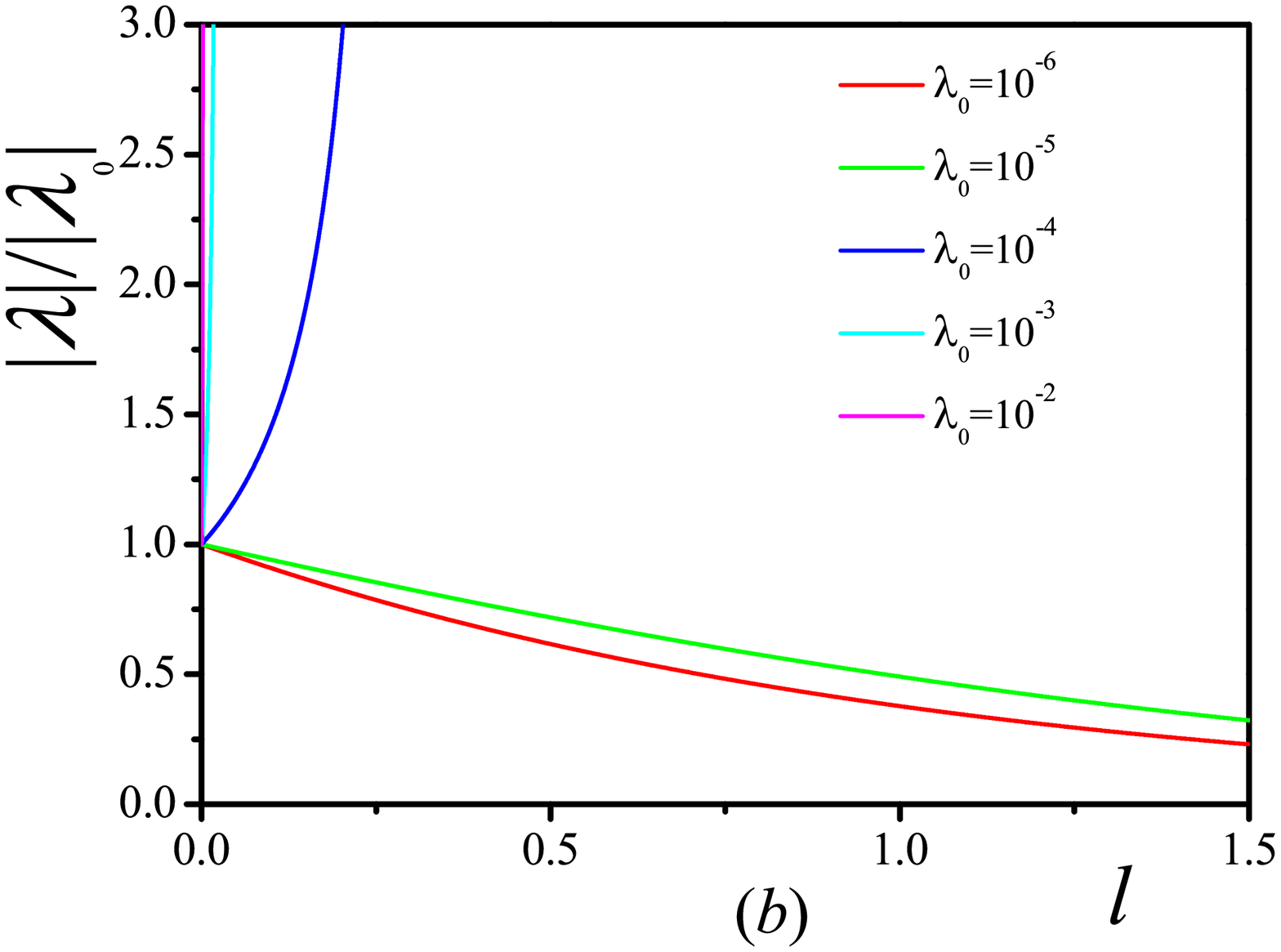}
\\ \vspace{-6.7cm}
\hspace{-0.1cm}\includegraphics[width=1.8in]{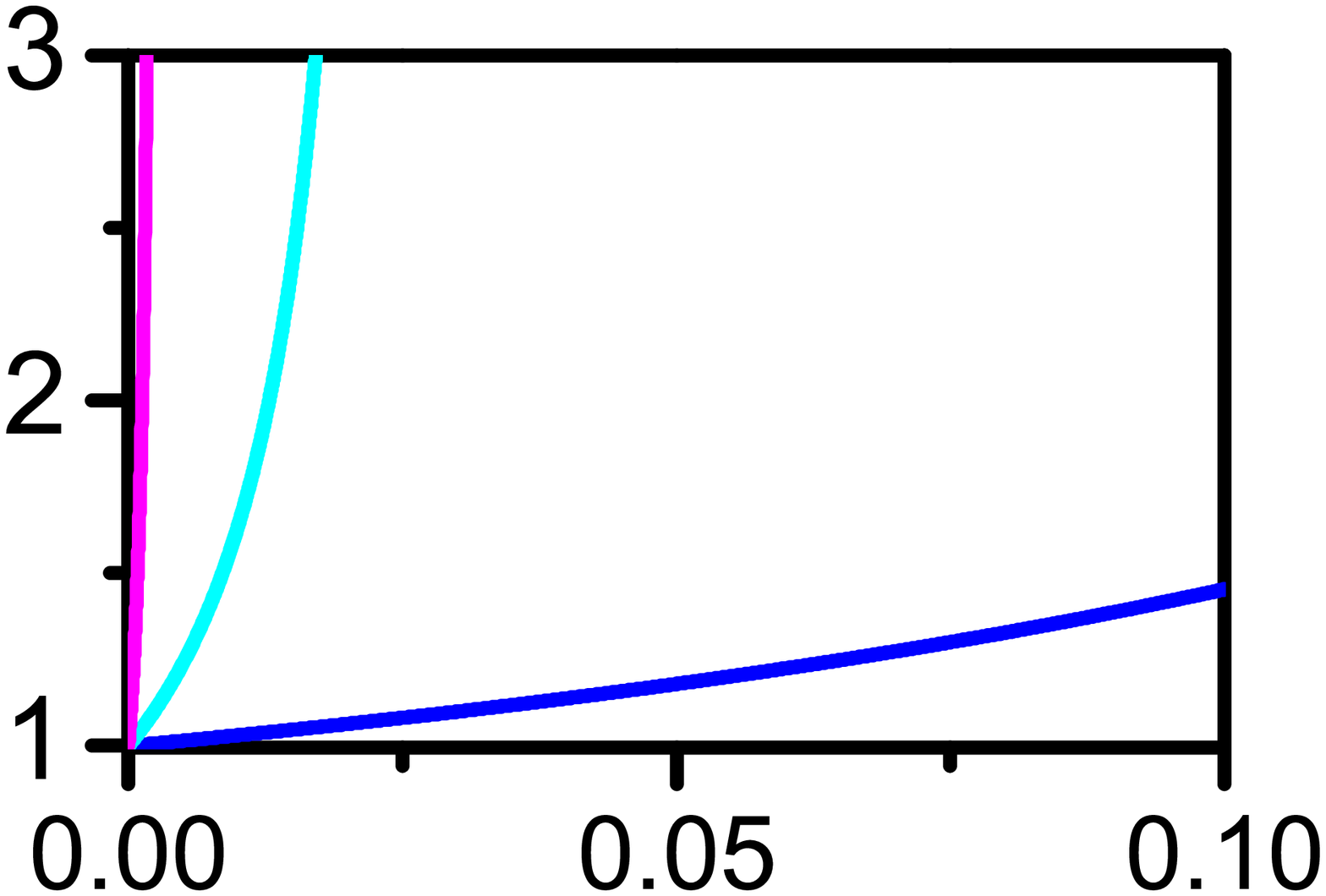}
\vspace{2.3cm}
\caption{(Color online) Evolutions of $|\lambda|$ upon lowering energy scales
for $\mu=0$ and $\alpha(0)=5\times10^{-3}$,  $v(0)=10^{-3}$: (a). $\lambda(0)=-10^{-4}$
and $\varphi=5\pi/6$ with several representative values of $Q$ and (b). $Q=10^{-3}$ and
$\varphi=\pi$ with several representative values of $\lambda_0\equiv\lambda(0)$. Insets: the enlarged
regions within the Cooper instability phases.}\label{Fig_mu0_lambda-Q_neq}
\end{figure}

To proceed, we initially endeavor to study $\lambda$'s evolution~(\ref{Eq_RG_lambda-2}) analytically.
One can infer the critical strength of starting value of $\lambda$ via
assuming Eq.~(\ref{Eq_RG_lambda-2})'s left hand side equals to zero, namely
\begin{eqnarray}
\lambda_c(0)&=&\frac{\pi^2}{\left(\sum^5_{i=3}\mathcal{D}_i-\mathcal{D}_2\right)}.\label{Eq_lambda_c-mu-0}
\end{eqnarray}
This forthrightly singles out that the Cooper instability can be formally ignited once
the initial strength $|\lambda(0)|$ exceeds the critical value $|\lambda_c(0)|$ while
the parameters $\mathcal{D}_i$ are regarded as constants.
However, it is of particular interest to point out that $\lim_{\mathbf{Q}\rightarrow0}
|\lambda_c(0)|\rightarrow\infty$ attesting to the defined
functions $\mathcal{D}_i(\mathbf{Q}=0)\rightarrow0$ with $i=2$ to $5$.
Therefore, the Cooper instability
is unable to be generated and this is consistent with our previous analyses that
the transfer momentum $\mathbf{Q}$ plays a crucial role. In order to explicitly show the
tendencies of parameter $\lambda$ upon decreasing the energy scale,
we are suggested to calculate the RG equation numerically by adopting
several representatively beginning values of correlated parameters.
Particularly, as the low-energy fate of $\lambda$ is closely linked
to the momentum $\mathbf{Q}$, we introduce two variables, i.e., $Q$ and
$\varphi$, to denote its strength and direction, respectively. As the
parameter $Q$ carries the property of momentum, we also measure it with the
cutoff $\Lambda_0$ in the following numerical calculations.
The corresponding results are gathered in Fig.~\ref{Fig_mu0_lambda-Q-pi_neq}
and Fig.~\ref{Fig_mu0_lambda-Q_neq}. We next address them in details.

At first, we make our focus on two special angles, at which the
$(\sum^5_{i=3}\mathcal{D}_i-\mathcal{D}_2)=0$
(independent of the value of $Q$), namely $\varphi_1=\pi/2$ and
$\varphi_2=3\pi/2$. Hence, they are equivalent to the case of $Q=0$ and
directly reduce to the tree level case (the lines are overlapped, namely independent of $Q$).
Therefore, the Cooper instability cannot be triggered as depicted in
Fig.~\ref{Fig_mu0_lambda-Q-pi_neq}(a) (the results for $\varphi_2=3\pi/2$
are the same to $\varphi_1=\pi/2$'s and thus are not shown in the figure).
For convenient reference, we hereby name these two special points
as $\varphi\in\mathrm{Zone-O}$.

Subsequently, all other angles cluster into two groups. We call them Zone-I and Zone-II
determined by $\alpha$, $v$, and $Q$,  at which $(\sum^5_{i=3}\mathcal{D}_i-\mathcal{D}_2)$
would be positive and negative respectively, namely
\begin{eqnarray}
\varphi\in\mathrm{Zone-I}:\,\,\,\left(\sum^5_{i=3}\mathcal{D}_i-\mathcal{D}_2\right)>0,\label{Eq_Zone-I}\\
\varphi\in\mathrm{Zone-II}:\,\,\,\left(\sum^5_{i=3}\mathcal{D}_i-\mathcal{D}_2\right)<0.\label{Eq_Zone-II}
\end{eqnarray}
We then consider them one by one. At Zone-I, we find that the Cooper instability
cannot be activated although it is sensitive to the transfer momentum $\mathbf{Q}$ upon increasing
$Q$ and $|\lambda_0|$ as shown in Fig.~\ref{Fig_mu0_lambda-Q-pi_neq}(b) for a representative
angle $\varphi=\pi/3$. To be concrete, this can be understood strictly. Compared to
the tree-level flow, it behaviors as $d\lambda/dl=-(1+C)\lambda$ with the constant $C>0$ at Zone-I,
which therefore cannot produce the Cooper instability. In a sharp contrast, the Cooper
instability can be generated at Zone-II with the same initial conditions of
Fig.~\ref{Fig_mu0_lambda-Q-pi_neq}. Choosing two representative angles $\varphi=0$
and $\varphi=\pi/3$ at Zone-II and carrying out the numerical evaluations give rise
to the results delineated in Fig.~\ref{Fig_mu0_lambda-Q_neq}. Studying from
Fig.~\ref{Fig_mu0_lambda-Q_neq}, we find the Cooper instability
can be manifestly triggered by virtue of
increasing $Q$ at a fixed $\lambda(0)$ delineated in Fig.~\ref{Fig_mu0_lambda-Q_neq}(a)
or enlarging $|\lambda(0)|$ at a fixed $Q$ illuminated in Fig.~\ref{Fig_mu0_lambda-Q_neq}(b)
for two representative Zone-II angles $\varphi=5\pi/6$ and $\varphi=\pi$, respectively.
It is worth pointing out that the basic results of Fig.~\ref{Fig_mu0_lambda-Q_neq}
are insensitive to the initial values of parameters, for instance $\alpha$ and
$v$ (we assume that they are small compared to the cutoff), which would
only determine the critical energy scale at which the Cooper instability sets in.
All these numerical results are in line with our above analytical analyses.

\begin{figure}
\centering
\includegraphics[width=4.5in]{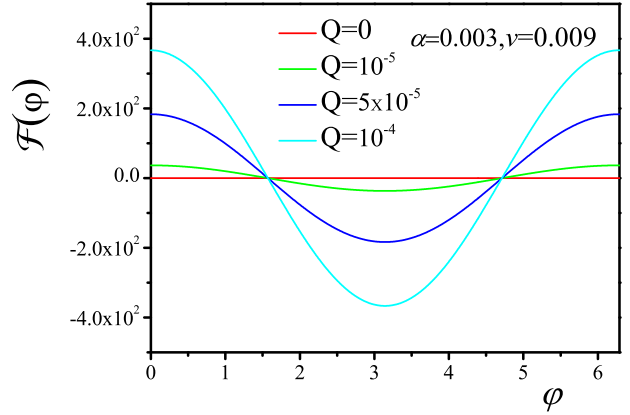}
\vspace{-2.3cm}
\caption{(Color online) The curves of $\mathcal{F}(\varphi)-\varphi$ at several representative
values of parameters $Q$, $v$, and $\alpha$ (the basic results are independent of these
values). }\label{Fig7_curves-F-varphi}
\end{figure}

Before going further, we hereby stop to present several discussions on
the ranges of both $\mathrm{Zone-I}$ and $\mathrm{Zone-II}$, which are
closely associated with the sign of $\left(\sum^5_{i=3}\mathcal{D}_i-\mathcal{D}_2\right)$.
To proceed, we nominate the $\mathcal{F}(\varphi)$ as
\begin{eqnarray}
\mathcal{F}(\varphi)&\equiv&\sum^5_{i=3}\mathcal{D}_i-\mathcal{D}_2=(\mathcal{M}\cos^3\varphi
+\mathcal{N}\cos^2\varphi\nonumber\\
&&+\mathcal{O}\cos\varphi+\mathcal{P})\cos\varphi.\label{Eq-f_varphi}
\end{eqnarray}
Then the slope of function $\mathcal{F}(\varphi)$ with respect to $\varphi$ can be derived as
\begin{eqnarray}
\mathcal{F}'(\varphi)&=&\frac{d\mathcal{F}(\varphi)}{d\varphi}=
-(4\mathcal{M}\cos^3\varphi+3\mathcal{N}\cos^2\varphi\nonumber\\
&&+2\mathcal{O}\cos\varphi+\mathcal{P})\sin\varphi,\label{Eq-f-prime_varphi}
\end{eqnarray}
where the coefficients $\mathcal{M}$, $\mathcal{N}$, $\mathcal{O}$, and
$\mathcal{P}$ are defined in Eqs.~(\ref{Eq-coeff_M})-(\ref{Eq-coeff_P}).


\begin{figure}
\hspace{-1.9cm}
\includegraphics[width=3.16in]{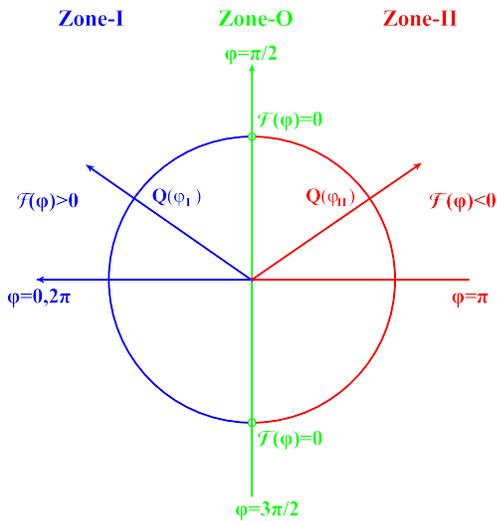}
\vspace{-0.0cm}
\caption{(Color online) Schematic descriptions for the ranges of
$\mathrm{Zone-O}$ ($\mathcal{F}(\varphi)=0$), and
$\mathrm{Zone-I}$ ($\mathcal{F}(\varphi)>0$) as well as
$\mathrm{Zone-II}$ ($\mathcal{F}(\varphi)<0$). $\mathbf{Q}_{\varphi_{\mathrm{I}}}$
and $\mathbf{Q}_{\varphi_{\mathrm{II}}}$ correspond to
two representatively transfer momenta in $Zone-I$ and
$Zone-II$, respectively. }\label{Fig8_F-varphi_physics-picture}
\end{figure}

To facilitate our discussions, we then divide the full region of $\varphi$
into Zone-O plus four subregions, namely $\mathrm{Region-A}\in[0,\frac{\pi}{2})$, $\mathrm{Region-B}\in(\frac{\pi}{2},\pi)$, $\mathrm{Region-C}\in(\pi,\frac{3\pi}{2})$, and
$\mathrm{Region-D}\in(\frac{3\pi}{2},2\pi]$. For $\varphi\in \mathrm{Region-A}$, we can straightforwardly
get $\mathcal{F}(\varphi=0)>0$ and $\lim_{\varphi\rightarrow\frac{\pi}{2}}\mathcal{F}(\varphi)=0$.
In addition, in order to make the two-dimensional semi-Dirac systems
stable, the distinction between the values of $\alpha$ and $v$ is very small
(if we assume $\alpha\gg v$ or $v\gg \alpha$, the
dispersion of our system would directly reduce to the full parabolic or linear situations).
Moveover, the value of transfer momenta $|\mathbf{Q}|$ is much smaller than other parameters.
Gather all these factors together, we can obtain that the slope of function $\mathcal{F}(\varphi)$
satisfies $\mathcal{F}'(\varphi)<0$. Therefore, $\mathcal{F}(\varphi)$ monotonically
decreases and gradually evolves from a positive value towards zero,
namely $\mathrm{Region-A}\in \mathrm{Zone-I}$.
Similarly, one can obtain $\mathrm{Region-D}$ also belongs to $\mathrm{Zone-I}$.
On the contrary, one can get $\lim_{\varphi\rightarrow\frac{\pi}{2}}\mathcal{F}(\varphi)=0$
and $\mathcal{F}'(\varphi)<0$ for $\varphi\in \mathrm{Region-B}$. Analogously,
for $\varphi\in \mathrm{Region-C}$,
it reads that $\mathcal{F}(\varphi=\pi)<0$ and
$\lim_{\varphi\rightarrow\frac{3\pi}{2}}\mathcal{F}(\varphi)=0$
as well as $\mathcal{F}'(\varphi)>0$.
Consequently, these analyses can tell us that $\mathcal{F}(\varphi)<0$
for $\mathrm{Region-B}$ and $\mathrm{Region-C}$,
namely, $\mathrm{Zone-II}$ consisting of both
$\mathrm{Region-B}$ and $\mathrm{Region-C}$.
Before closing the discussions, we stress
again that the sign of function $\mathcal{F}$ strongly depends upon
$\varphi$ although its value is slightly sensitive to $Q$.
Based on both analytical and numerical analysis above,
one can find that the overall structures and signs of $\mathcal{F}(\varphi)$ are
very insensitive to $Q$ but are even solely determined by $\varphi$ as long as
$Q$ is small (this can be always satisfied as $Q$ is a transfer momentum)
as manifestly shown in Fig.~\ref{Fig7_curves-F-varphi} and
Fig.~\ref{Fig8_F-varphi_physics-picture}.

In order to verify our analytical discussions, we perform the numerical
calculations via taking several representative values of $Q$, $v$, and $\alpha$
and obtain the the results shown in Fig.~\ref{Fig7_curves-F-varphi}, which are well
consistent with above analyses. Based on above discussions, we can draw a conclusion
that the Cooper instability can be generated while $\varphi$ is restricted to $\mathrm{Zone-II}$
with a nonzero value of $|\mathbf{Q}|$. Additionally,
we would like to stress that $\mathrm{Zone-II}$ including
both $\mathrm{Region-B}$ and $\mathrm{Region-C}$ is not small but nearly takes half
of full directions (regions) ($\frac{\pi}{2}$ and $\frac{3\pi}{2}$ do not belong to $\mathrm{Zone-II}$)
as schematically shown in Fig.~\ref{Fig8_F-varphi_physics-picture}.
Before closing this section, let us hereby address some comments on possibly
physical pictures for the ranges of $\mathrm{Zone-I}$ and
$\mathrm{Zone-II}$. Specifically, if we consider the regions $\varphi\in(\frac{\pi}{2},\frac{3\pi}{2})$ and
$\varphi\in[0,\frac{\pi}{2}]\mathrm{U}[\frac{3\pi}{2},2\pi]$ as the so
called ``\emph{forward-alike scattering}"
and ``\emph{back-alike scattering}", respectively, one may realize that above analysis implies
that only the``\emph{forward-alike scattering}" associated with
the transfer momentum $\mathbf{Q}_{\varphi_{\mathrm{II}}}$ can contribute to the Cooper instability as
schematically shown in Fig.~\ref{Fig8_F-varphi_physics-picture}.
On the contrary, the ``\emph{back-alike scattering}"
can not provide useful corrections to Cooper instability. This picture
should be also physically reasonable.

To recapitulate, we have examined how the Cooper-pairing interaction influences
the low-energy states of 2D SD materials at $\mu=0$, in particular the possibility of
Cooper instability. Table~\ref{table_Cooper_phase} and Fig.~\ref{Fig8_F-varphi_physics-picture}
summarize our main results for both $\mu=0$ and $\mu\neq0$. In next subsection, we are going
to investigate the situation in the presence of a finite chemical potential.

\begin{table}[t]
\caption{Collections of basic conclusions for Cooper instability (CI) due to one-loop corrections
of Cooper-pairing interaction for both zero and a finite chemical potential.
The terminology ``\emph{CI always generated}" means that the CI can be
triggered at an arbitrarily weak Cooper-pairing coupling strength $\lambda$.
The ``\emph{Zone-O}" corresponds to $\varphi=\frac{\pi}{2},\frac{3\pi}{2}$ and
``\emph{Zone-I}" and ``\emph{Zone-II}" are designated in Eq.~(\ref{Eq_Zone-I})
and Eq.~(\ref{Eq_Zone-II}), respectively.\label{table_Cooper_phase}}
\vspace{0.39cm}
\centerline{
\begin{tabular}{ l |l} 
\hline
\hline
$\mu=0$, $Q=0$ or $\varphi\in\mathrm{Zone-O}$  & No CI  \\
\hline
$\mu=0$, $Q\neq0$, $\varphi\in \mathrm{Zone-I}$  & No CI  \\
\hline
$\mu=0$, $Q\neq0$, $\varphi\in \mathrm{Zone-II}$  &
CI triggered at $|\lambda(0)|>|\lambda_c(0)|$  \\
\hline
$\mu\neq0$ & CI always generated \\
\hline
\hline
\end{tabular}
}
\end{table}

\subsection{$\mu$-tuned phase transition}

As addressed at the beginning of this section, the $\mu$-tuned phase transition
is expected in that the DOS at Fermi surface for zero chemical potential is qualitatively distinct
from a finite-$\mu$ situation's~\cite{Neto2009RMP,Banerjee2009PRL,Banerjee2012PRB}.
Under such circumstance, one naturally concerns the question whether and how this
phase transition is linked to the Cooper instability.

\begin{figure}
\centering
\includegraphics[width=4.6in]{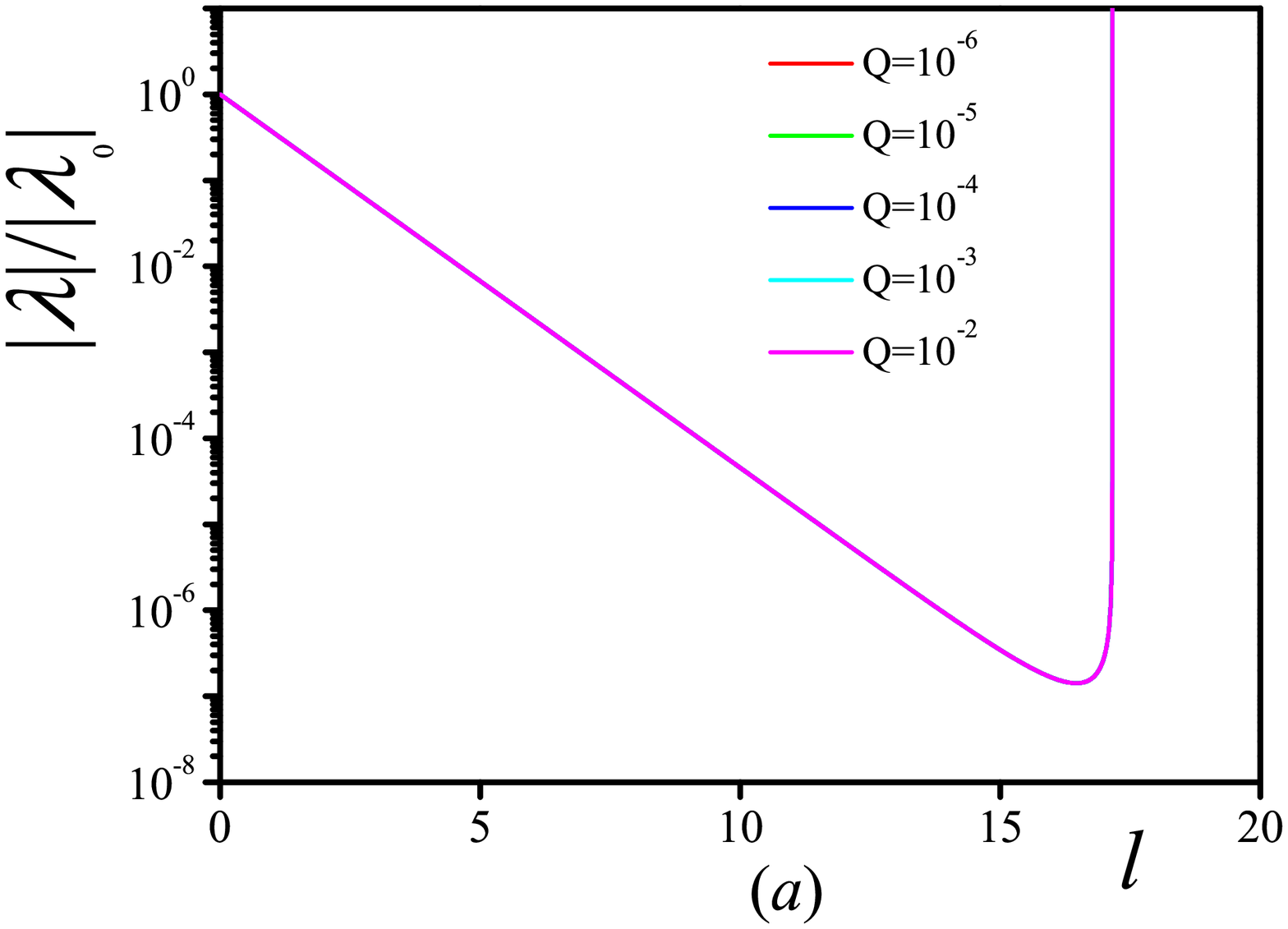}
\\ \vspace{-5.35cm}
\hspace{-1.2cm}\includegraphics[width=1.8in]{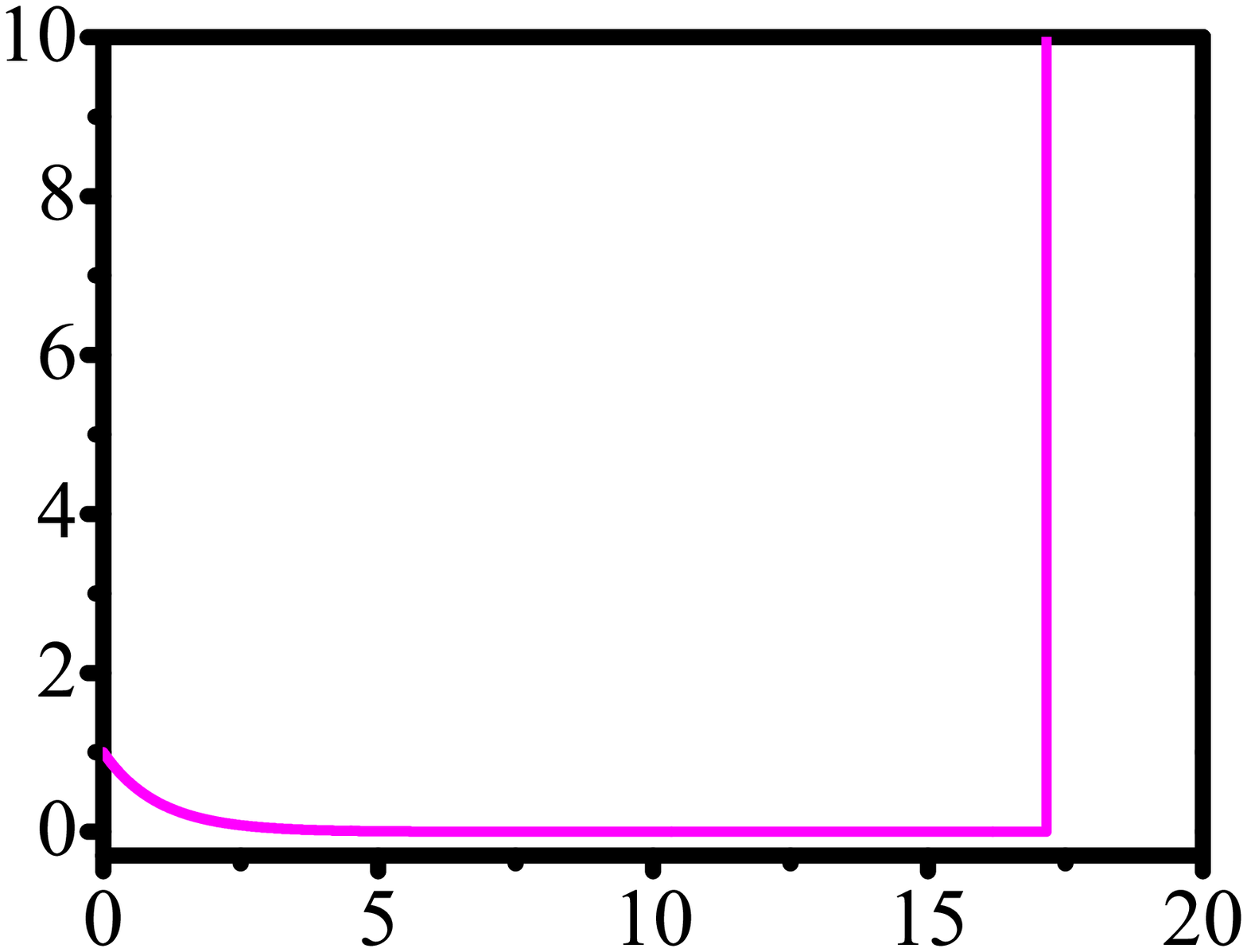}
\\ \vspace{0.16cm}
\includegraphics[width=4.6in]{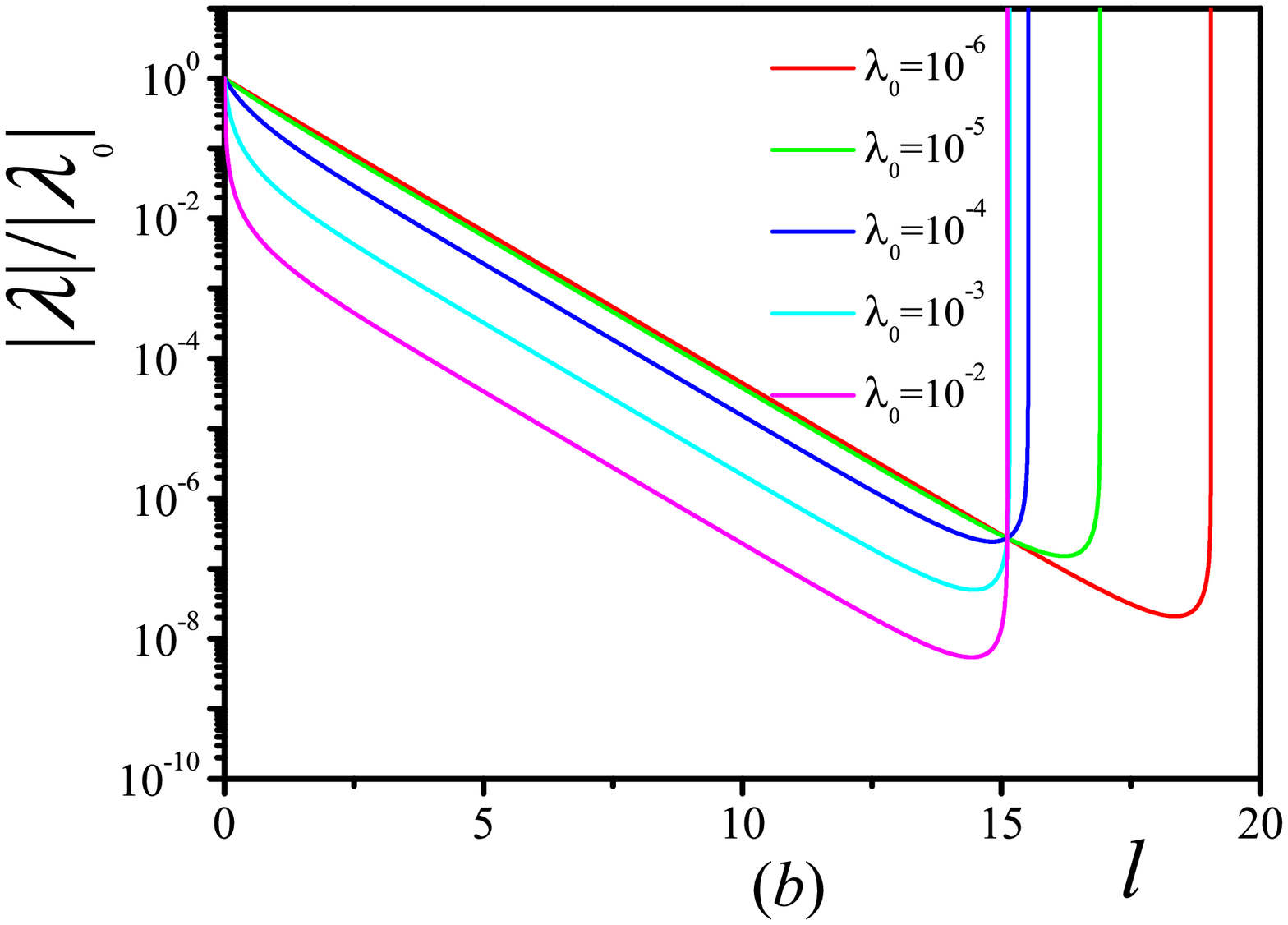}
\\\vspace{-5.35cm}
\hspace{-1.1cm}\includegraphics[width=1.8in]{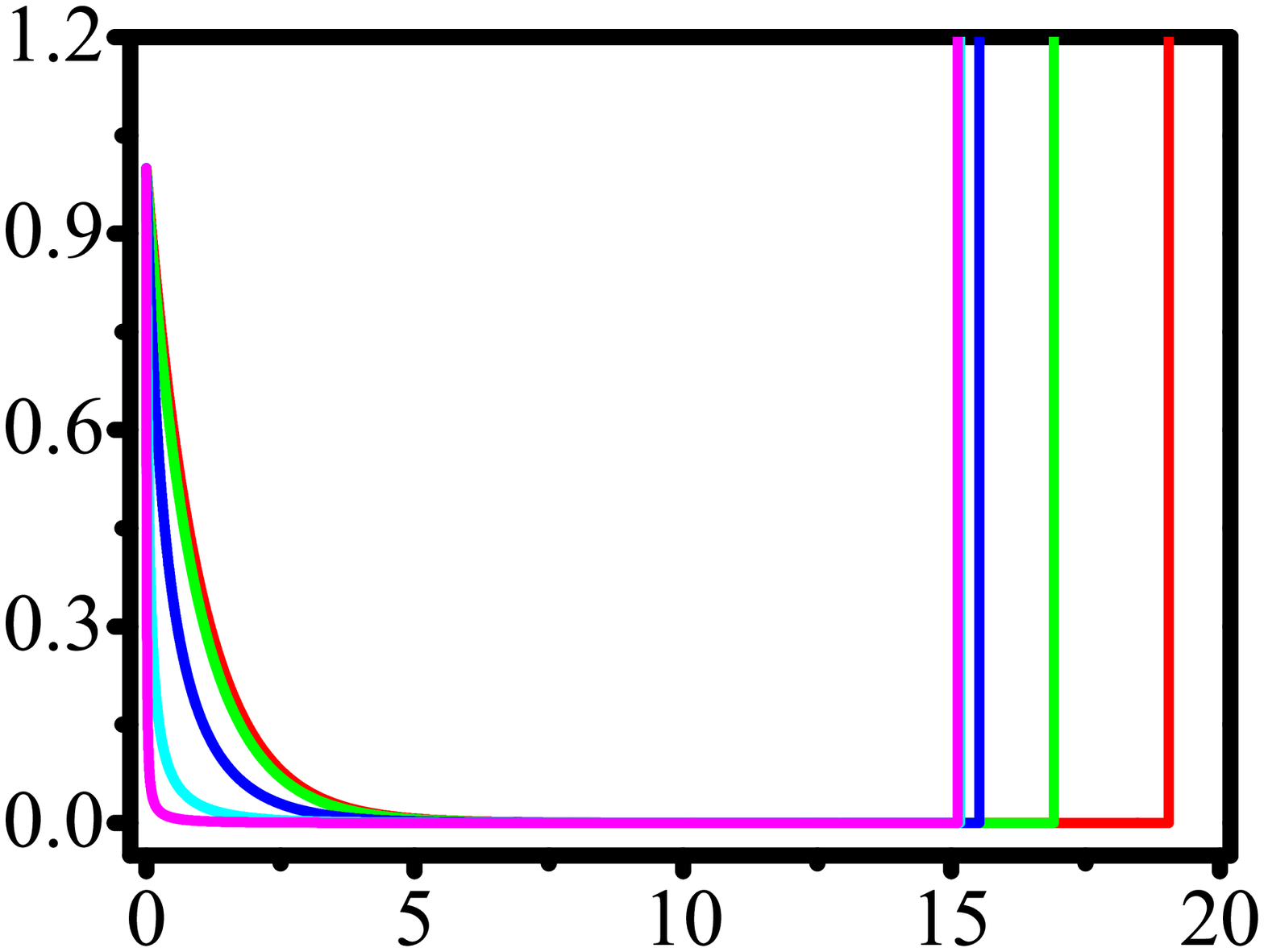}
\vspace{0.3cm}
\caption{(Color online) Evolutions of $|\lambda|$ upon lowering energy scales
for $\mu(0)=10^{-5}$ and $\alpha(0)=5\times10^{-3}$,  $v(0)=10^{-3}$:
(a). $\lambda(0)=-10^{-4}$ and $\varphi=\pi/2$
with several representative values of $Q$ and (b). $Q=10^{-3}$ and
$\varphi=\pi/3$ with several representative values of $\lambda_0\equiv\lambda(0)$.
Insets: the additional version of these curves are produced by adopting the same data
to directly make comparisons with their counterparts in Fig.~\ref{Fig_mu0_lambda-Q-pi_neq}
(the qualitative tendencies are independent of the
specific values of the chemical potential).} \label{Fig_mu_lambda-Q_neq}
\end{figure}

To response these, paralleling the analysis for $\mu=0$ part,
we can initially derive the formal $\lambda_c(0)$ with hypothesizing all other
parameters to be constants by virtue of referring to Eq.~(\ref{Eq_RG_lambda}),
\begin{eqnarray}
\lambda_c(0)&=&\frac{4\pi^2}{4\left(\sum^5_{i=3}\mathcal{D}_i-\mathcal{D}_2\right)
-\mu^2\mathcal{D}_0}.\label{Eq_lambda_c-mu-finite}
\end{eqnarray}
Before going further, we recall pieces of useful information obtained in
Sec.~\ref{Sec_mu0}: $D_i(Q\rightarrow0)=0$ or $D_i(\varphi=\pi/2)=D_i(\varphi=3\pi/2)=0$
with $i=2$ to $5$ and the sign of $(\sum^5_{i=3}\mathcal{D}_i-\mathcal{D}_2)$ is
positive or negative respectively corresponding to $\varphi\in\mathrm{Zone-I}$
and $\varphi\in\mathrm{Zone-II}$. With respect to this information,
this critical coupling, at the first sight, is very analogous
to the case with $\mu=0$, $Q\neq0$, and $\varphi\in \mathrm{Zone-II}$, indicating
the Cooper instability being produced at $|\lambda(0)|>|\lambda_c(0)|$ as
listed in Table~\ref{table_Cooper_phase}. However, we would like to
emphasize that these two circumstances are qualitatively distinct. In the former,
the coupling $\lambda_c(0)$ are constants that determined by the values of
$\alpha$, $v$, and $Q$. Conversely, the $\lambda_c(0)$ for the latter evolves towards
zero in that the chemical potential $\mu$ is a relevant quantity
by means of RG term as characterized in Eq.~(\ref{Eq_RG_mu}), which
climbs up upon lowering the energy scales. As a result, it implies any weak attractive
interaction can ignite the Cooper pairing once a finite $\mu$ is introduced,
namely the Cooper theorem~\cite{BCS1957PR}. This result is well consistent
with the mean-field analysis of 2D Dirac semimetals~\cite{Uchoa-Neto2005PRB,Kopnin2008PRL},
which can be generally understood as follows. As a finite $\mu$ changes the Dirac
point and the DOS is nonzero at Fermi surface, this causes the BCS diagram
also contributes to the parameter $\lambda$, which becomes the very dominant subchannel.
To explicitly display the process, the numerical evolutions of $\lambda$ for the
presence of a representative $\mu$ are provided in Fig.~\ref{Fig_mu_lambda-Q_neq}(a) at $D_i=0$.
To proceed, an intriguing question is raised whether the outcome above is
sufficiently robust against a finite $Q$ at Zone-II, namely the fate of competition between $4(\sum^5_{i=3}\mathcal{D}_i-\mathcal{D}_2)$ and $\mu$. In order to response this,
we would like to select out several representatively starting values of parameters at Zone-II,
which are the same to their counterparts in Fig.~\ref{Fig_mu0_lambda-Q_neq}.
Additionally, we bring out a very small starting value of $\mu$, for instance $\mu=10^{-5}$
and numerically evaluate the running
evolutions of $\mu$ and $\lambda$~(\ref{Eq_RG_mu})-(\ref{Eq_RG_lambda}), leading to the
corresponding results in Fig.~\ref{Fig_mu_lambda-Q_neq}(b). To reiterate, we
stress that the basic results of Fig.~\ref{Fig_mu_lambda-Q_neq} are
insensitive to the concrete beginning values of $\mu$.

Reading off the information in Fig.~\ref{Fig_mu_lambda-Q_neq} and gathering all these analyses and
discussions together, we therefore come to a conclusion that a finite $\mu$ indeed play an
essential role in triggering the Cooper instability and a $\mu$-tuned phase transition associated with
the Cooper instability can be expected~\cite{Sachdev1999Book,Vojta2003RPP}.

\section{Cooper instability influenced by the impurity scattering at $\mu=0$}\label{Sec_imp}

It is well known that the impurities are present in nearly all fermionic systems,
whose effects on the low-energy behaviors of physical quantities are widely investigated~\cite{Ramakrishnan1985RMP,Nersesyan1995NPB,Mirlin2008RMP,
Efremov2011PRB,Efremov2013NJP,Korshunov2014PRB,Fiete2016PRB,
Nandkishore2013PRB,Potirniche2014PRB,Nandkishore2017PRB,Stauber2005PRB,Roy2016PRB-2,
Roy1604.01390,Roy1610.08973,Roy2016SR,Aleiner2006PRL,Aleiner2006PRL-2,
Montambaux-Orignac2016PRB,Lee1702.02742}.
Generally, impurity scatterings can induce the damping rate of fermions,
which can both promote fermionic excitations with shortening their lifetimes to be harmful
for the superconductivity and enhance the DOS of fermionic systems
to be helpful for the superconductivity.
Accordingly, it is worth asking how the impurity influences
the Cooper instability under the competition between these two adverse sorts of contributions.

As addressed in previous section, we attentively
investigate the emergence of Cooper instability
for the presence of both zero and a finite chemical potential at clean limit. One of most significant
points in this situation is that a finite chemical potential $\mu$ plays a central role
in low-energy regime and can always induce the Cooper instability. Therefore,
we here focus on the situation at $\mu=0$ and briefly discuss the effects of impurities
on the formation of Cooper instability.

Based on one-loop corrections provided in Appendix~{\ref{Appendix_impurity-calculations}},
we arrive at the updated RG equations of interaction parameters for the presence of
multi types of impurities,
\begin{eqnarray}
\frac{d\alpha}{dl}&=&
\frac{[-(\Delta_C+\Delta_{G_1})\mathcal{E}_0]\alpha}{4\pi^2},\label{Eq_alpha-3}\\
\frac{dv}{dl}&=&\frac{[-(\Delta_C+\Delta_M)\mathcal{E}_0]v}{4\pi^2},\label{Eq_v-3}\\
\frac{d\Delta_C}{dl}&=&\frac{\Bigl[\sum_{I\neq C}\Delta_I\mathcal{E}_1
-\Delta_C\mathcal{E}_2\Bigr]\Delta_C}{4\pi^2},\label{Eq_C-3}\\
\frac{d\Delta_{G_1}}{dl}&=&\frac{1}{4\pi^2}\Bigl[-(\Delta_{G_3}+\Delta_M )\mathcal{E}_1
-\Delta_{G_1} \mathcal{E}_2\nonumber\\
&&+\Delta_C (\mathcal{E}_1-2\mathcal{E}_2)\Bigr]\Delta_{G_1},\label{Eq_G1-3}\\
\frac{d\Delta_{G_3}}{dl}&=&\frac{1}{4\pi^2}\Bigl[(\Delta_{G_1}+\Delta_{M})\mathcal{E}_1
-\Delta_{G_3}\mathcal{E}_2\nonumber\\
&&+\Delta_C(\mathcal{E}_1-2\mathcal{E}_0)\Bigr]\Delta_{G_3},\label{Eq_G3-3}\\
\frac{d\Delta_M}{dl}&=&\frac{1}{4\pi^2}\Bigl[(\Delta_{G_1}+\Delta_{G_3})
(\mathcal{E}_1-2\mathcal{E}_2)-\Delta_C\mathcal{E}_1\nonumber\\
&&-\Delta_M\mathcal{E}_2+4\lambda(\mathcal{E}_1-\mathcal{E}_2)\Bigr]\Delta_M,\label{Eq_M-3}
\end{eqnarray}
combined together with
\begin{eqnarray}
\frac{d\lambda}{dl}\!=\!\left[\!-1-\frac{\lambda\left(4\mathcal{D}_2-4\sum^5_{i=3}\mathcal{D}_i\right)}
{4\pi^2}-\frac{\sum_I\Delta_I \mathcal{E}_2}{4\pi^2}\!\right]\!\!\lambda.\label{Eq_lambda-3}
\end{eqnarray}
where the parameters again $C$, $G_{1,3}$, and $M$ denote the random chemical potential,
random gauge potential and random mass, respectively.

\begin{figure}
\centering
\includegraphics[width=4.6in]{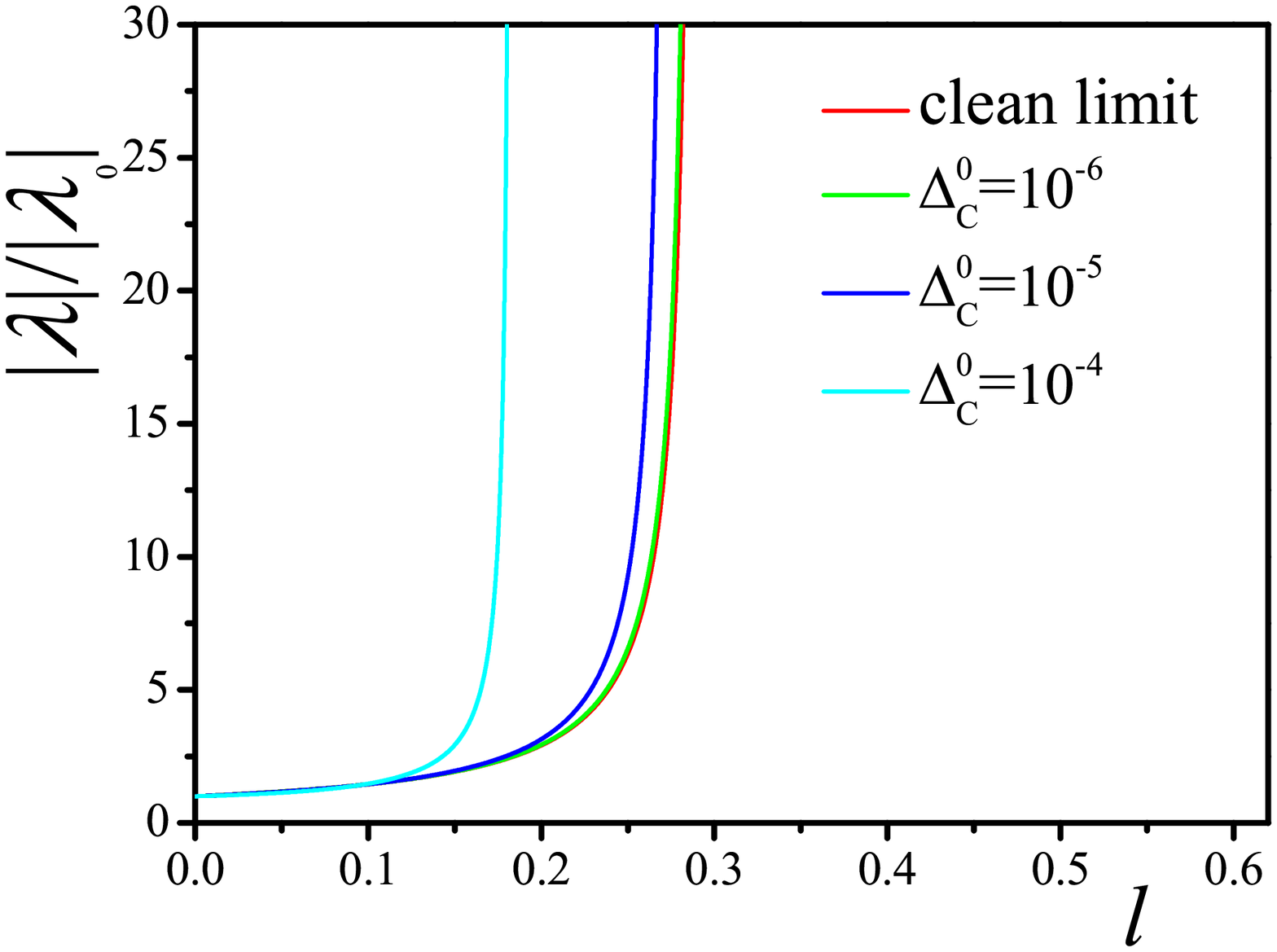}
\\ \vspace{-5.39cm}
\hspace{3.96cm}\includegraphics[width=1.36in]{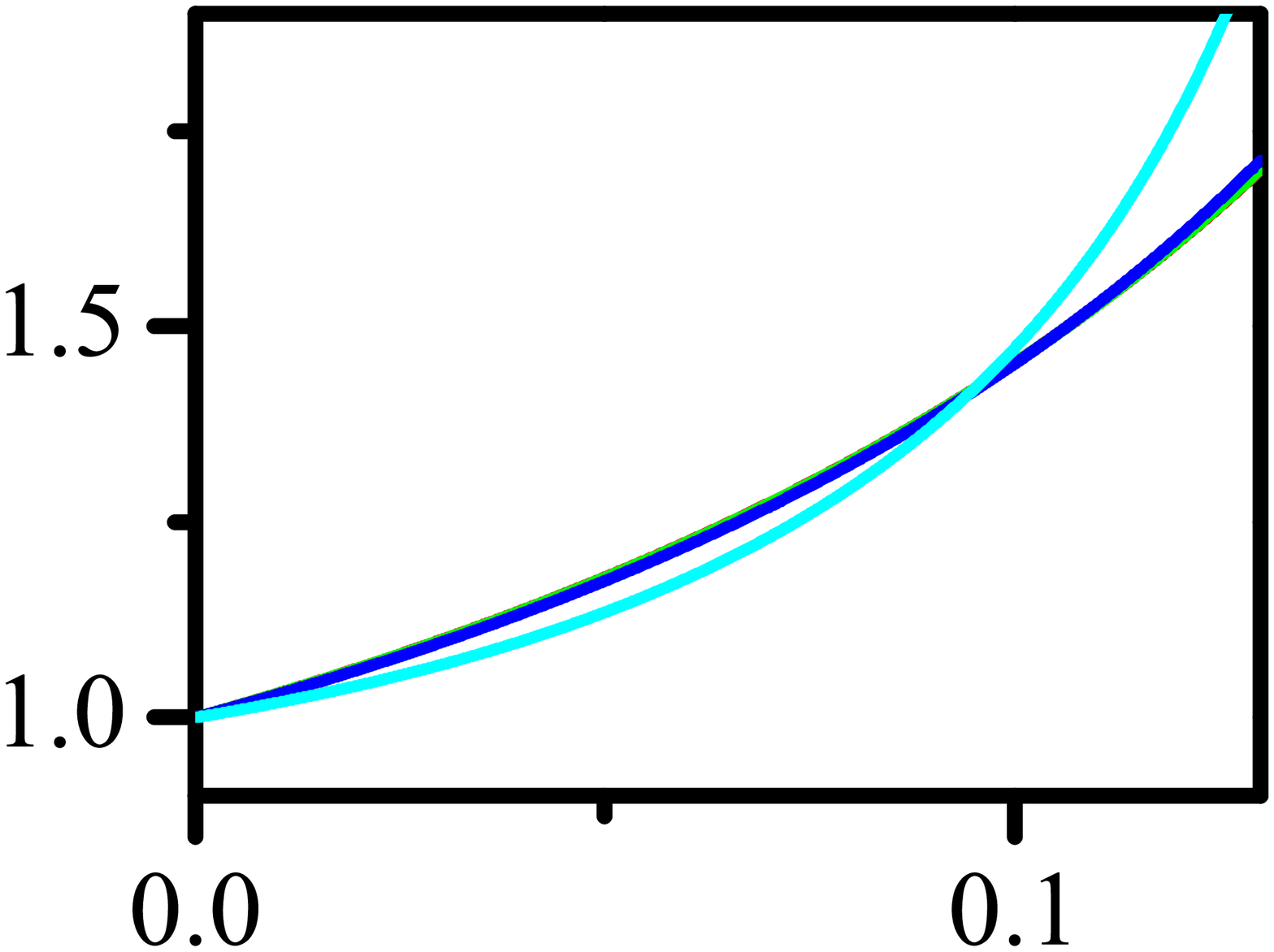}
\vspace{0.8cm}
\caption{(Color online) Comparisons of energy-dependent
evolutions of $|\lambda|$ between the clean limit and presence of random chemical potential
for typical initial values of impurity strengths with several representatively initial values
of parameters, i.e., $\mu=0$, $\alpha(0)=5\times10^{-3}$,  $v(0)=10^{-3}$, $Q=10^{-3}$,
$\varphi=\pi$ and $\lambda_0=-10^{-4}$ (We would like to
emphasize that the impurity strength is usually weak in real systems and thus from now on
the large strengths are introduced only for theoretical exhibitions). Inset: the enlarged regions for the starts of evolutions.}\label{Fig10_C_contrast-lambda-M4}
\end{figure}

\begin{figure}
\centering
\includegraphics[width=4.6in]{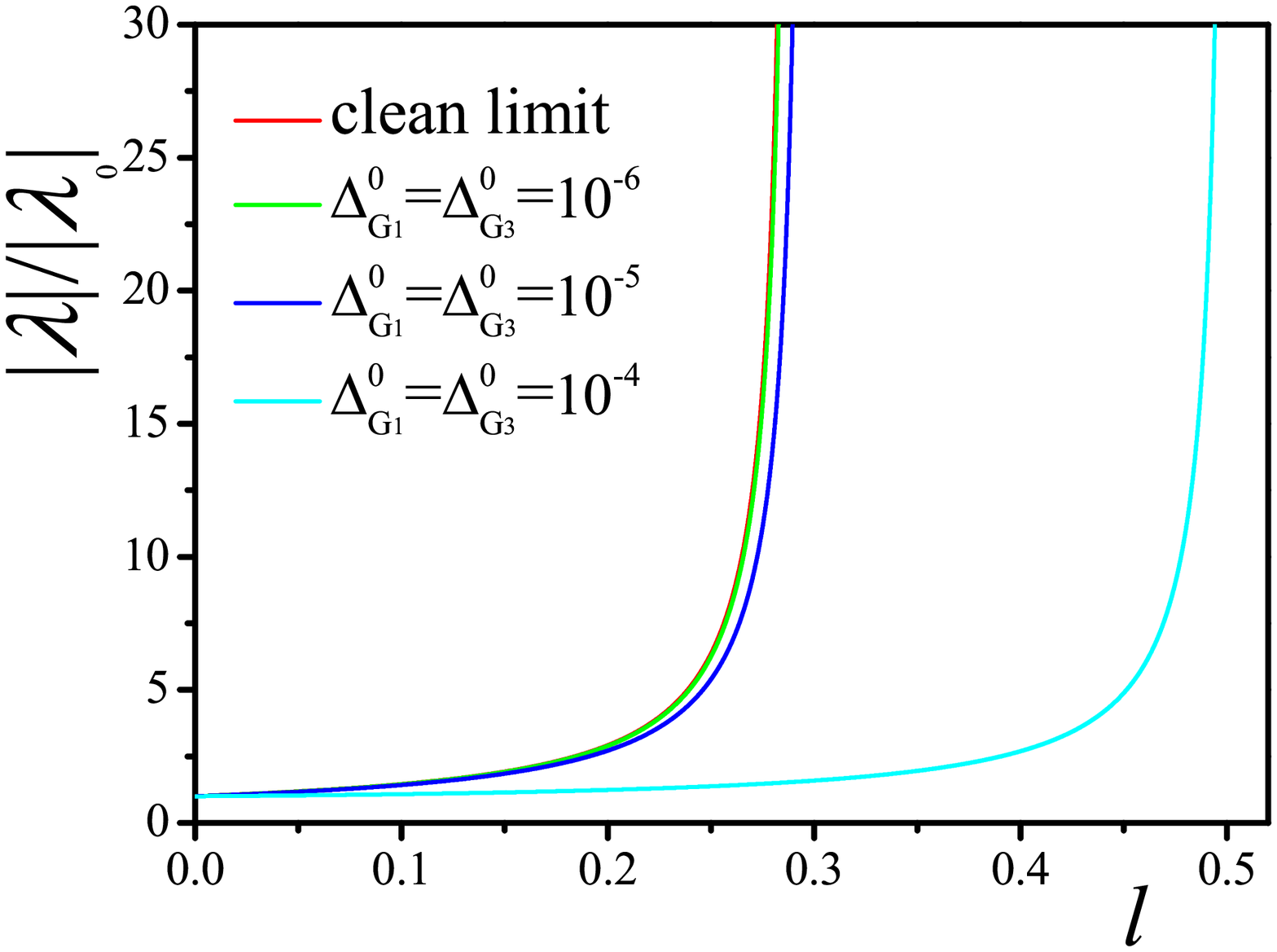}
\\ \vspace{-5.39cm}
\hspace{-3.0cm}\includegraphics[width=1.36in]{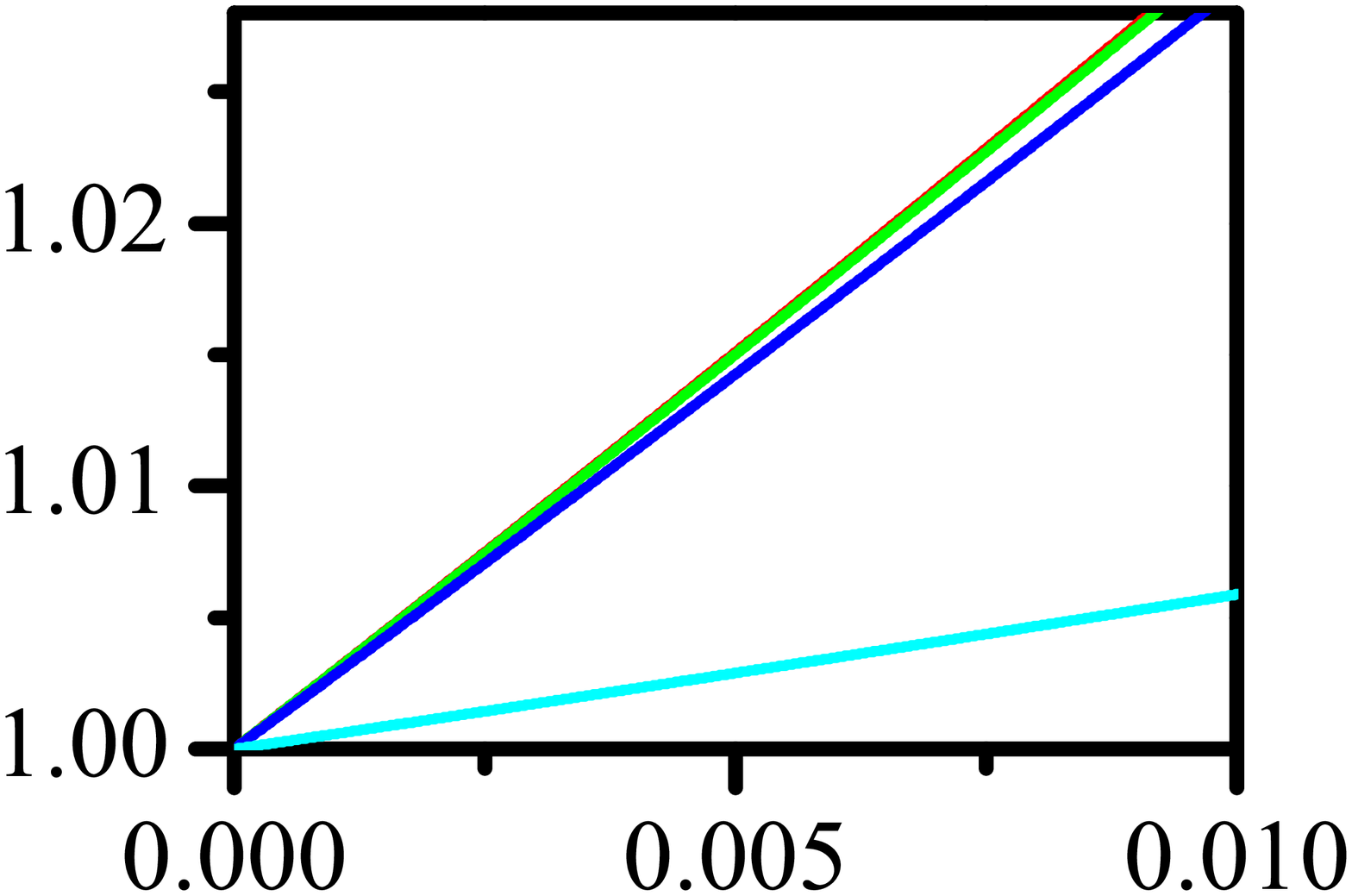}
\vspace{0.8cm}
\caption{(Color online) Comparisons of energy-dependent
evolutions of $|\lambda|$ between the clean limit and presence of random gauge potential
for several initial values of impurity strengths with several representatively initial values
of parameters, i.e., $\mu=0$, $\alpha(0)=5\times10^{-3}$,  $v(0)=10^{-3}$, $Q=10^{-3}$,
$\varphi=\pi$ and $\lambda_0=-10^{-4}$.
Inset: the enlarged regions for the starts of evolutions.}\label{Fig11_G1G3_contrast-lambda-M4}
\end{figure}
\begin{figure}
\centering
\includegraphics[width=4.6in]{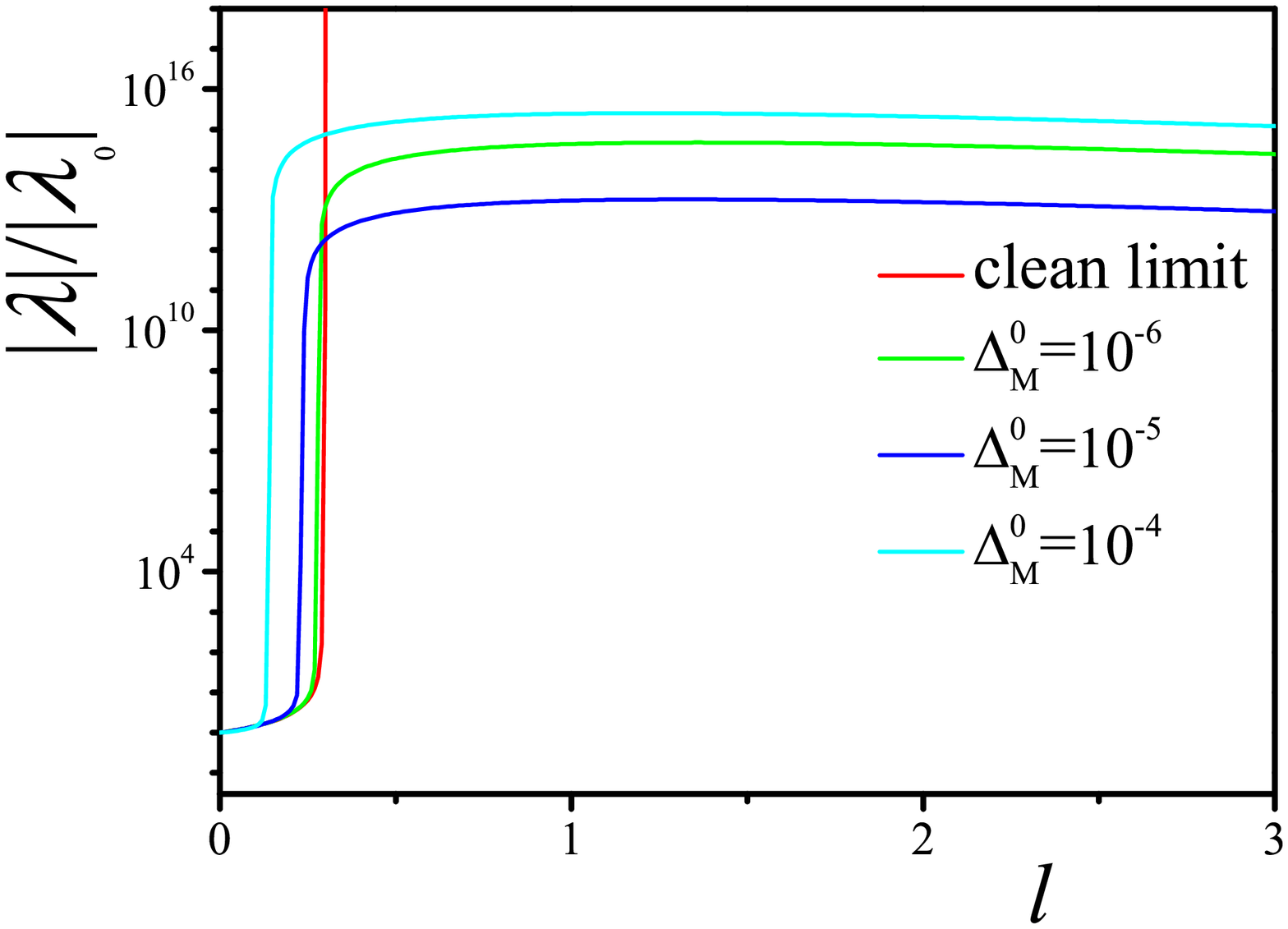}
\\ \vspace{-5.6cm}
\hspace{-0.3cm}\includegraphics[width=1.56in]{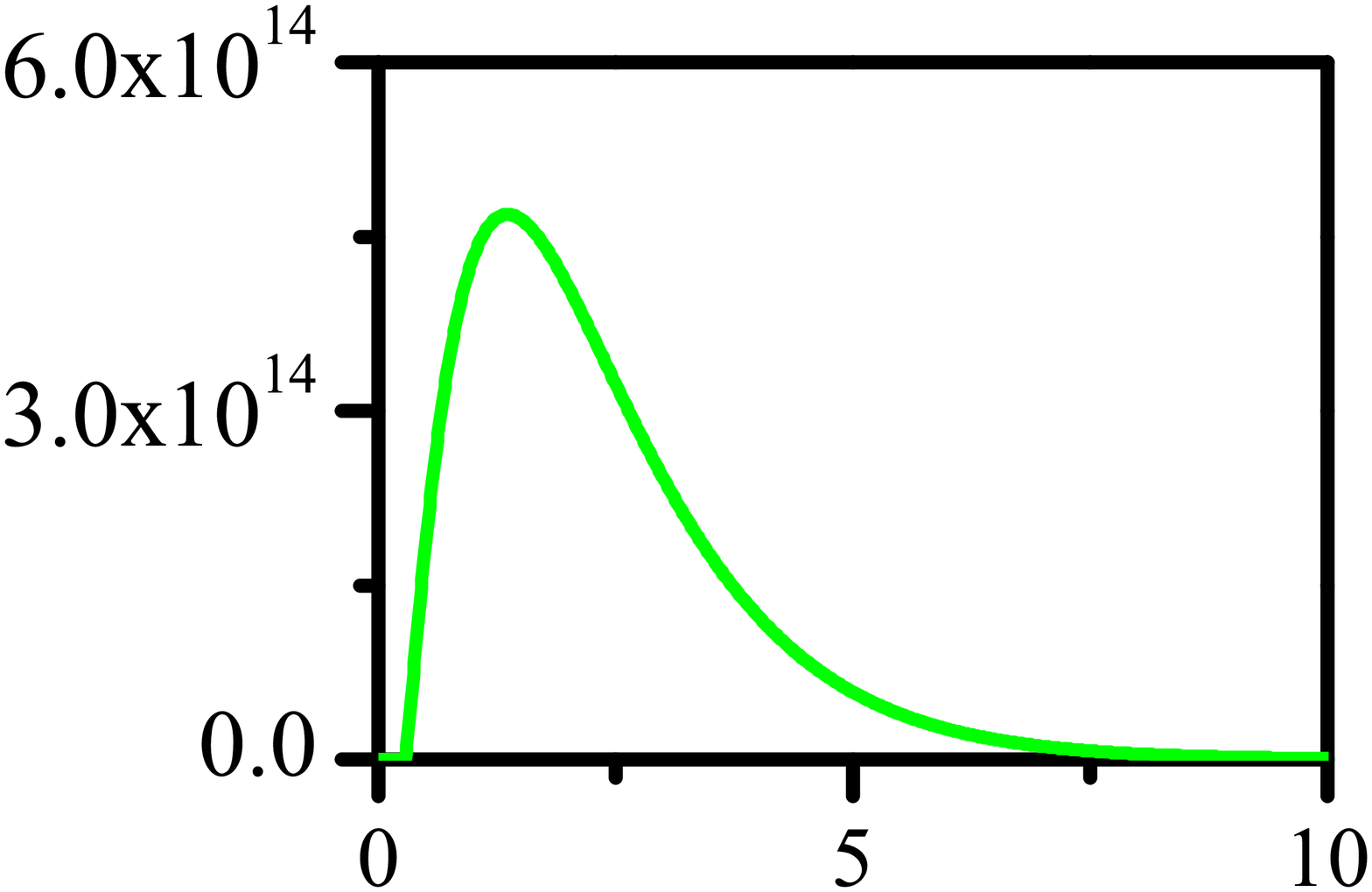}
\vspace{1.0cm}
\caption{(Color online) Comparisons of energy-dependent
evolutions of $|\lambda|$ between the clean limit and presence of random mass
for several initial values of impurity strengths with several representatively initial values
of parameters, i.e., $\mu=0$, $\alpha(0)=5\times10^{-3}$,  $v(0)=10^{-3}$, $Q=10^{-3}$,
$\varphi=\pi$ and $\lambda_0=-10^{-4}$. Inset: the enlarged regions of $\lambda(l)$
for $\Delta^0_M=10^{-6}$ (the basic results are similar for other initial
values of impurity strengths and hence not shown here). }\label{Fig12_Mass_contrast-lambda-M4}
\end{figure}

Subsequently, we examine the influence of impurity on the Cooper instability.
At first, we can derive the formally critical value of $\lambda$ via paralleling the
analyses in Sec.~\ref{Sec_mu0}. Reading off Eq.~(\ref{Eq_lambda-3}),
the $\lambda_c(0)$ can be extracted as,
\begin{eqnarray}
\lambda_c(0)&=&\frac{\pi^2\left(1+\frac{\sum_I\Delta_I \mathcal{E}_2}{4\pi^2}\right)}
{\left(\sum^5_{i=3}\mathcal{D}_i-\mathcal{D}_2\right)}.\label{Eq_lambda_c-mu-0-2}
\end{eqnarray}
In comparison with its clean-limit and $\mu=0$ counterpart~(\ref{Eq_lambda_c-mu-0}),
one can directly find that the critical strength
$\lambda_c(0)$ is manifestly increased due to the impurity scatterings
as long as the parameters $\alpha$ and $v$ are constants.
However, we need to emphasize that Eqs.~(\ref{Eq_alpha-3})-(\ref{Eq_lambda-3})
unambiguously exhibit that all parameters are not independent but intimately
coupled with each others. With this respect, it is necessary to perform the numerical
calculation along these coupled flow equations to explicitly show the
fate of parameter $\lambda$ at low-energy scales.

As it is of particular interest to examine the effects of impurity scatterings on
the Cooper instability, we hereby only focus on the
situations at which the Cooper instability
can be ignited potentially at the clean limit, namely $\varphi\in\mathrm{Zone-II}$.
For completeness, the presence of single and multi sorts of
impurities will be both investigated. At the outset, we assume
only one type of impurity exists in the system. To proceed, we derive
the reduced RG equations for the presence of random chemical potential
via assuming $\Delta_{I\neq C}=0$ in Eqs.~(\ref{Eq_alpha-3})-(\ref{Eq_lambda-3}),
\begin{eqnarray}
\frac{d\alpha}{dl}&=&\frac{-\Delta_C\mathcal{E}_0}{4\pi^2}\alpha,\label{Eq_alpha-33}\\
\frac{dv}{dl}&=&\frac{-\Delta_C\mathcal{E}_0}{4\pi^2}v,\label{Eq_v-33}\\
\frac{d\Delta_C}{dl}
&=&\frac{-\Delta^2_C\mathcal{E}_2}{4\pi^2},\label{Eq_C-33}\\
\frac{d\lambda}{dl}&=&\left(-1-\lambda_\mathcal{D}-\frac{\Delta_C \mathcal{E}_2}{4\pi^2}\right)\lambda,\label{Eq_lambda-34}
\end{eqnarray}
with designating $\lambda_\mathcal{D}\equiv\lambda\left(\mathcal{D}_2-\sum^5_{i=3}\mathcal{D}_i\right)/(\pi^2)$.

One can obviously read that the random chemical potential is decreased progressively upon
lowering the energy scales, namely an irrelevant quantity in the RG terminology. This indicates
that its effects are gradually weakened as the energy is lowered due to its irrelevant feature.
However, the $\lambda$ couples with the strength of impurity as well as
the parameters $\alpha$ and $v$, which also evolve and
are associated with the $\mathcal{D}_i$ and $\lambda_{\mathcal{D}}$. This implies
that the value of $\lambda_c(0)$~(\ref{Eq_lambda_c-mu-0-2}) can either be increased or
lowered caused by the energy-dependent evolutions of $\mathcal{D}_i$. In order to judge this,
we therefore need to perform the numerical
analyses of these reduced evolutions of Eqs.~(\ref{Eq_alpha-33})-(\ref{Eq_lambda-34}).
To straightforwardly compared with the clean-limit case, we start with the same starting conditions
of Fig.~\ref{Fig_mu0_lambda-Q_neq}, which contains the main results at $\mu=0$.
To be specific, we assign two typical values to $Q$ and $\varphi$, for instance
$Q=10^{-3}$ and $\varphi=\pi$ as utilized in Fig.~\ref{Fig_mu0_lambda-Q_neq}(b).
After performing numerical calculations of
Eqs.~(\ref{Eq_alpha-33})-(\ref{Eq_lambda-34}), we find that the Cooper instability is
fairly insensitive to this sort of impurity, which only results in the increase of the
critical energy scale (i.e., decrease of $l_c$) where the instability is ignited
as designated in Fig.~\ref{Fig10_C_contrast-lambda-M4}. In other words, attesting to
the evolutions of parameters $\alpha$ and $v$, the single presence of random
chemical potential is slightly favorable to the Cooper instability although
the impurity is an irrelevant quantity. This is of particular distinction from the Dirac semimetals~\cite{Sondhi2013PRB,Sondhi2014PRB,Wang2017PRB_BCS}.

\begin{figure}
\centering
\includegraphics[width=4.6in]{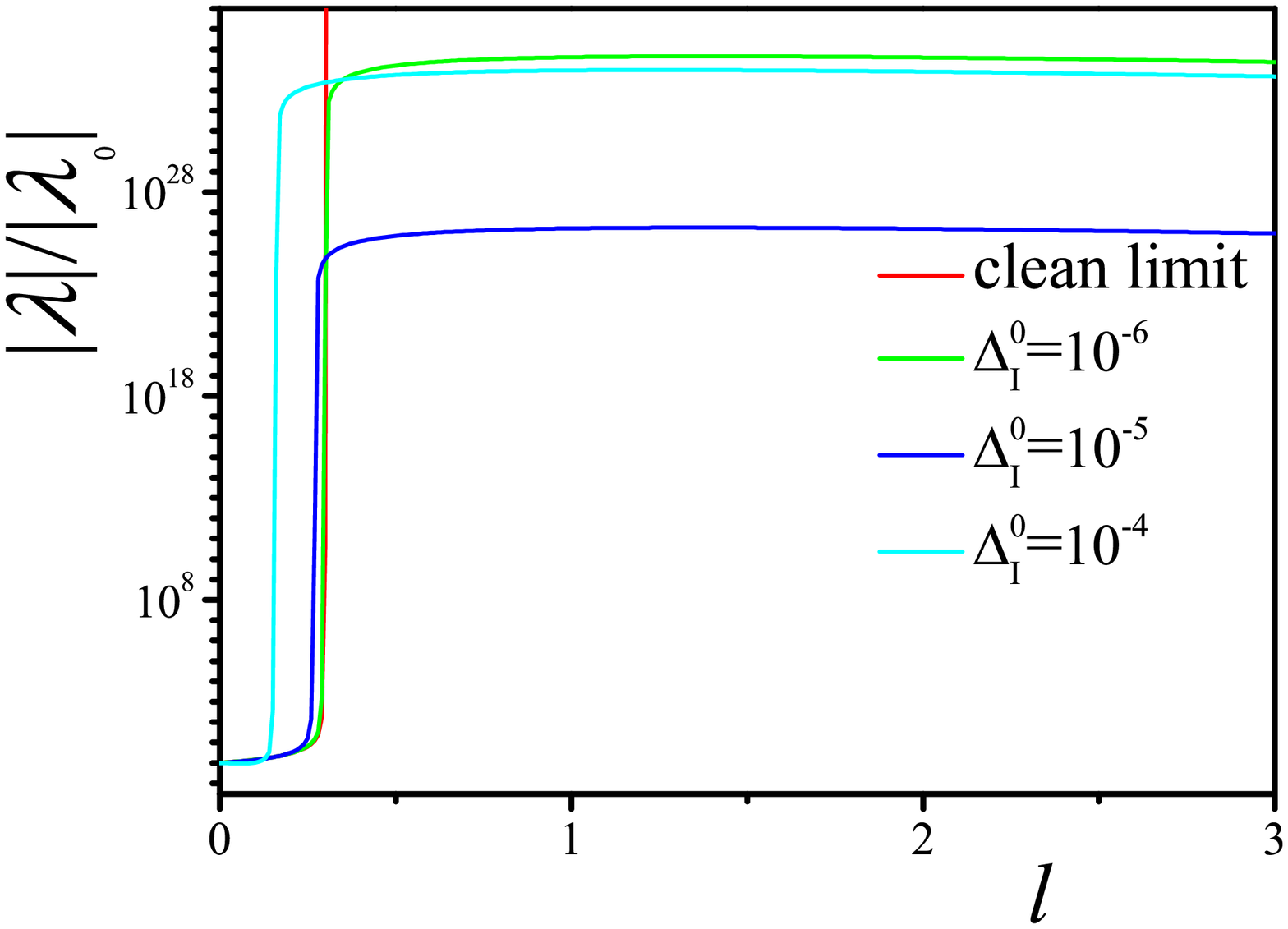}
\\ \vspace{-5.6cm}
\hspace{-0.3cm}\includegraphics[width=1.56in]{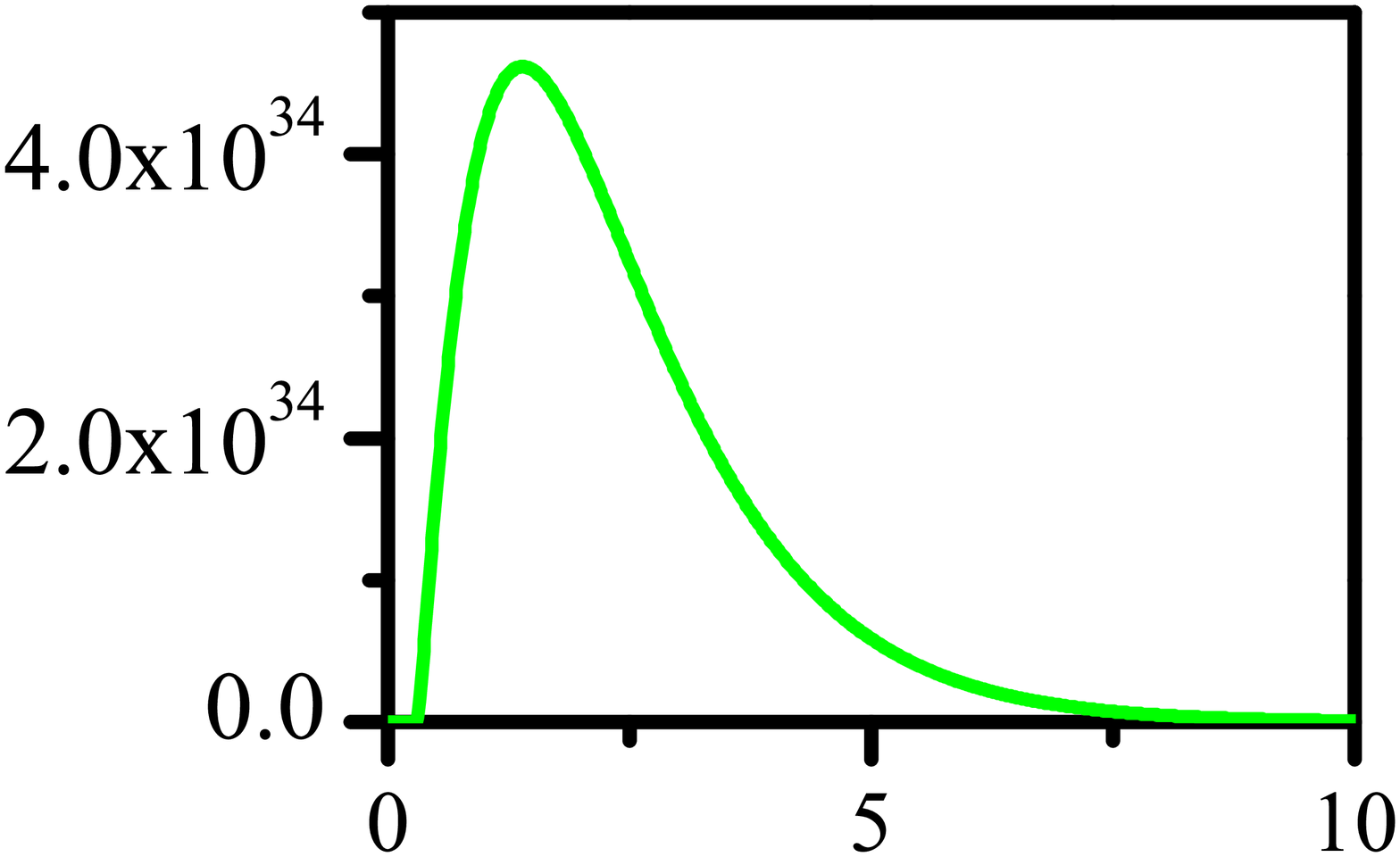}
\vspace{1.0cm}
\caption{(Color online) Comparisons of energy-dependent
evolutions of $|\lambda|$ between the clean limit and presence of all three sorts of impurities
for several initial values of impurity strengths with several representatively initial values
of parameters, i.e., $\mu=0$, $\alpha(0)=5\times10^{-3}$,  $v(0)=10^{-3}$, $Q=10^{-3}$,
$\varphi=\pi$ and $\lambda_0=-10^{-4}$. Without loss of generality, all three types
of impurities are assumed to own the equally starting strengths, namely
$\Delta_{C}=\Delta_{G_{1,3}}=\Delta_{M}=\Delta_{I}$. Inset: the enlarged regions of $\lambda(l)$
for $\Delta^0_I=10^{-6}$ (the basic results are similar for other initial
values of impurity strengths and hence not shown here).}\label{Fig13_CGM_contrast-lambda-M4}
\end{figure}

In addition, for the presence of only random gauge potential or random mass,
the coupled evolutions reduce to
\begin{eqnarray}
\frac{d\alpha}{dl}&=&
\frac{-\Delta_{G_1}\mathcal{E}_0}{4\pi^2}\alpha,\,\,\frac{dv}{dl}=0,\label{Eq_v-5}\\
\frac{d\Delta_{G_1}}{dl}&=&\frac{(-\Delta_{G_3}\mathcal{E}_1-\Delta_{G_1} \mathcal{E}_2)}{4\pi^2}\Delta_{G_1},\label{Eq_G1-5}\\
\frac{d\Delta_{G_3}}{dl}&=&\frac{(\Delta_{G_1}\mathcal{E}_1
-\Delta_{G_3}\mathcal{E}_2]}{4\pi^2}\Delta_{G_3},\label{Eq_G5}\\
\frac{d\lambda}{dl}&=&\left[-1-\lambda_\mathcal{D}-\frac{(\Delta_{G_1}+\Delta_{G_3}) \mathcal{E}_2}{4\pi^2}\right]\lambda,\label{Eq_lambda-5}
\end{eqnarray}
or
\begin{eqnarray}
\frac{d\alpha}{dl}&=&0,\,\,\frac{dv}{dl}=\frac{-\Delta_M\mathcal{E}_0}{4\pi^2}v,\label{Eq_v-6}\\
\frac{d\Delta_M}{dl}&=&\frac{4\lambda(\mathcal{E}_1-\mathcal{E}_2)-\Delta_M\mathcal{E}_2}
{4\pi^2}\Delta_M,\label{Eq_M-6}\\
\frac{d\lambda}{dl}&=&\left(-1-\lambda_\mathcal{D}-\frac{\Delta_M \mathcal{E}_2}{4\pi^2}\right)\lambda.\label{Eq_lambda-6}
\end{eqnarray}
To proceed, after paralleling the analyses for the single presence of random
chemical potential and carrying out analogous numerical analyses along the Eqs.~(\ref{Eq_v-5})-(\ref{Eq_lambda-6}), we obtain the primary the results for the
random gauge potential and random mass as delineated
in Fig.~\ref{Fig11_G1G3_contrast-lambda-M4} and Fig.~\ref{Fig12_Mass_contrast-lambda-M4},
respectively. To be concrete, we find that the critical energy scales are decreased
(i.e., increase of $l_c$) by the influence of random gauge potential as depicted in Fig.~\ref{Fig11_G1G3_contrast-lambda-M4}.
As a result, in distinction to the random chemical potential, the sole presence
of random gauge potential is slightly harmful to the Cooper instability. In a
sharp contrast, the coupled evolution equations for the sole presence of random
mass yield to several unusual features compared to both random chemical potential
and random gauge potential as designated in Fig.~\ref{Fig12_Mass_contrast-lambda-M4}.
At first, it is of particular importance to address that the critical value of
fermion-fermion strength $|\lambda(l)|$, as manifestly
illustrated in Fig.~\ref{Fig13_CGM_contrast-lambda-M4}, quickly climbs certain saturate peak
and then gradually flows toward zero at the lowest energy limit. In addition, the
starting points of saturate lines are shifted to bigger energy scales with the
increase of impurity's initial values. In other words, the fermion-fermion interaction's strength
$|\lambda(l)|$ is no longer divergent. As a consequence, the Cooper instability
is switched off by the random mass with an adequately large starting value.

Next, we consider the situation for presence of all three sorts of impurities.
Without loss of generality, all these three types of impurities are assumed to own the equally
starting strengths. To proceed, paralleling the analogous steps combined with the coupled flow
equations~(\ref{Eq_alpha-3})-(\ref{Eq_lambda-3}) leads to the main results
shown in Fig.~\ref{Fig13_CGM_contrast-lambda-M4}. According to this figure,
we find the fate of fermion-fermion strength is similar to the sole presence of
random mass. However, it is of particular significance to point out that
the saturated values are much higher due to the competition among
these three sorts of impurities. Accordingly, distinct types of impurities
compete with each other and eventually the random mass becomes dominant. In other words,
this indicates that the fermionic excitations promoted by impurities play a more important
role than the increase of DOS in 2D SD semimetals. To reiterate, we hereby
stress that the competition among distinct sorts of impurities is harmful to the Cooper
instability no matter whether any of single types of
impurities promotes or hinders the Cooper instability.

In brief, it is well known that there is a long history for the the effect
of impurity on the superconductivity~\cite{Anderson1959JPCS,Gorkov2008Book,
Balatsky2006RMP,Ramakrishnan1985RMP}, which is a complicate problem and
attracted a number of studies for both conventional and unconventional superconductors.
We admit that our focuses here are only on the qualitative roles of three types
of impurities and the studies here are somehow tentative. Despite this, one can expect the
results would provide several significant signals and tendencies
of these impurities in the low-energy regime.

\begin{figure}
\centering
\includegraphics[width=4.6in]{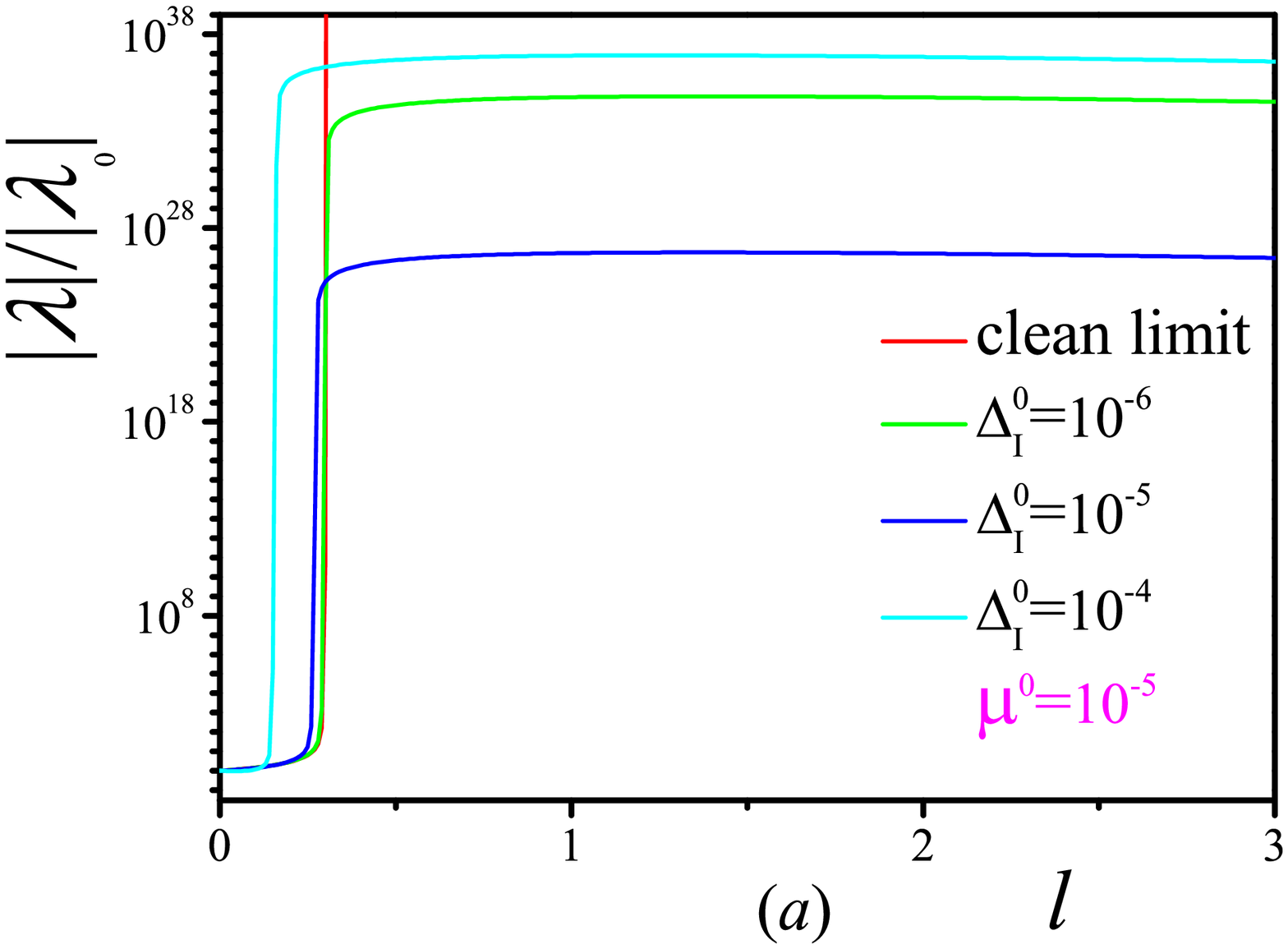}
\\ \vspace{-5.2cm}
\hspace{-0.6cm}\includegraphics[width=1.56in]{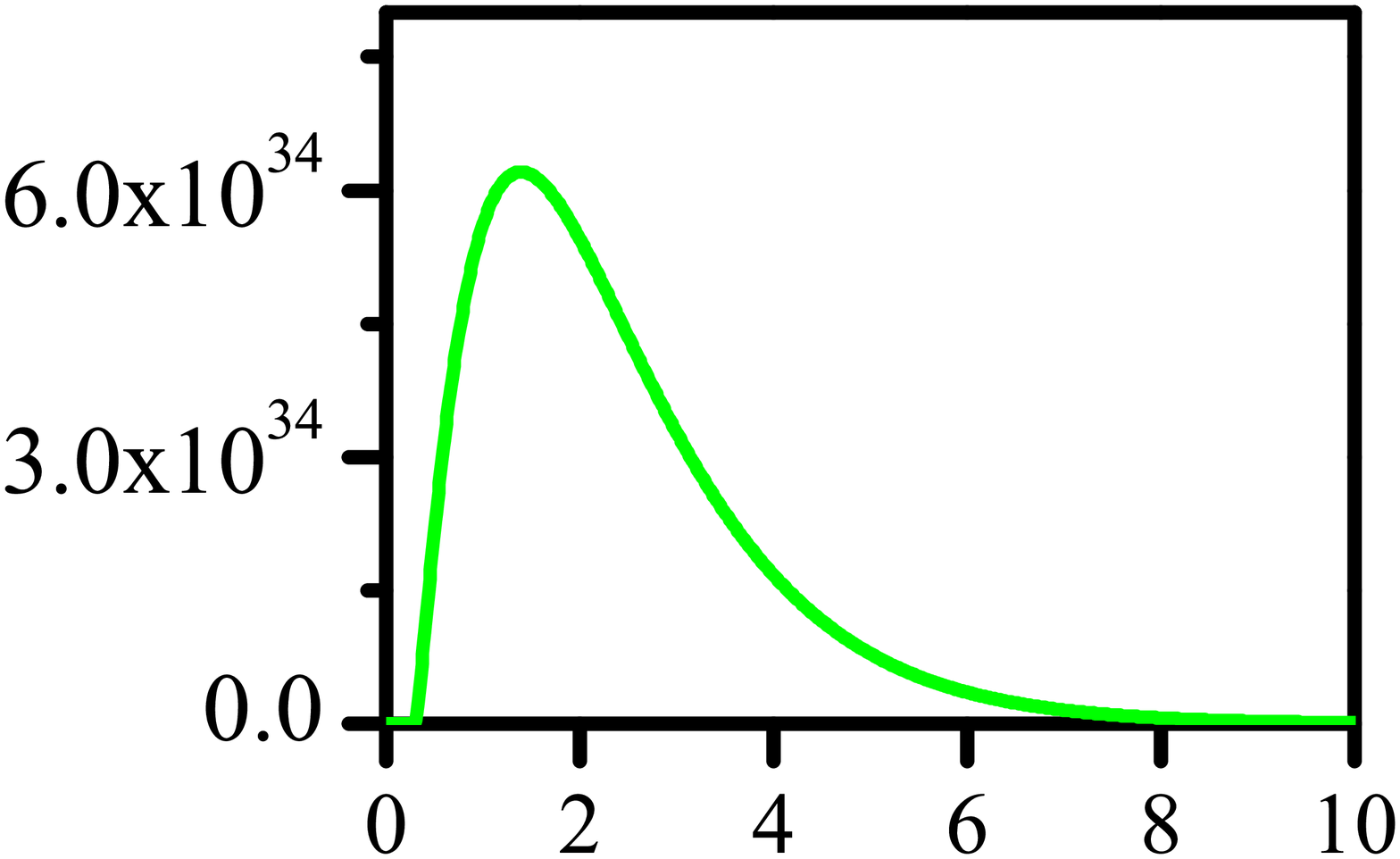}\\ \vspace{0.5cm}
\includegraphics[width=4.6in]{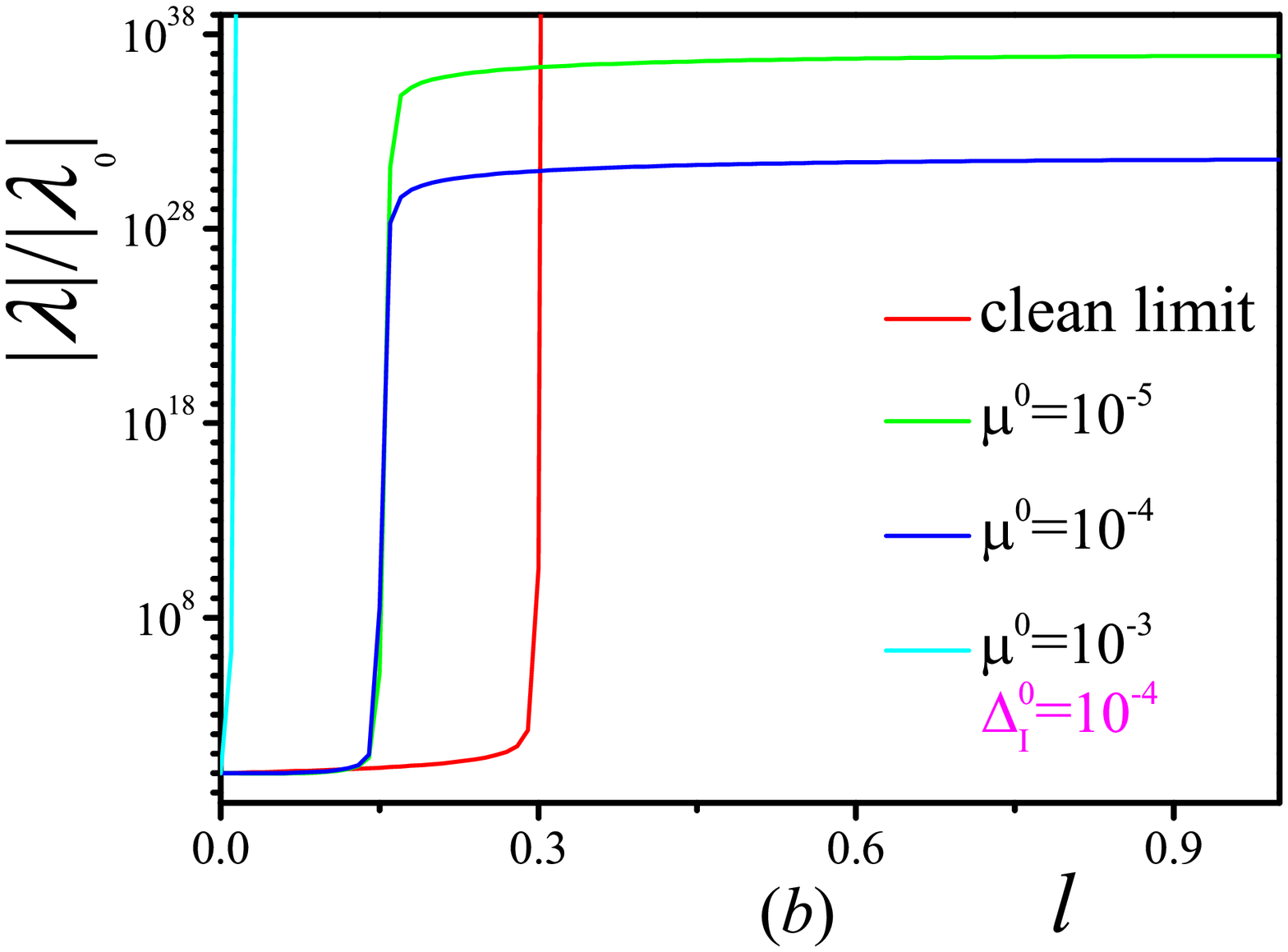}
\\ \vspace{-5.2cm}
\hspace{-0.6cm}\includegraphics[width=1.56in]{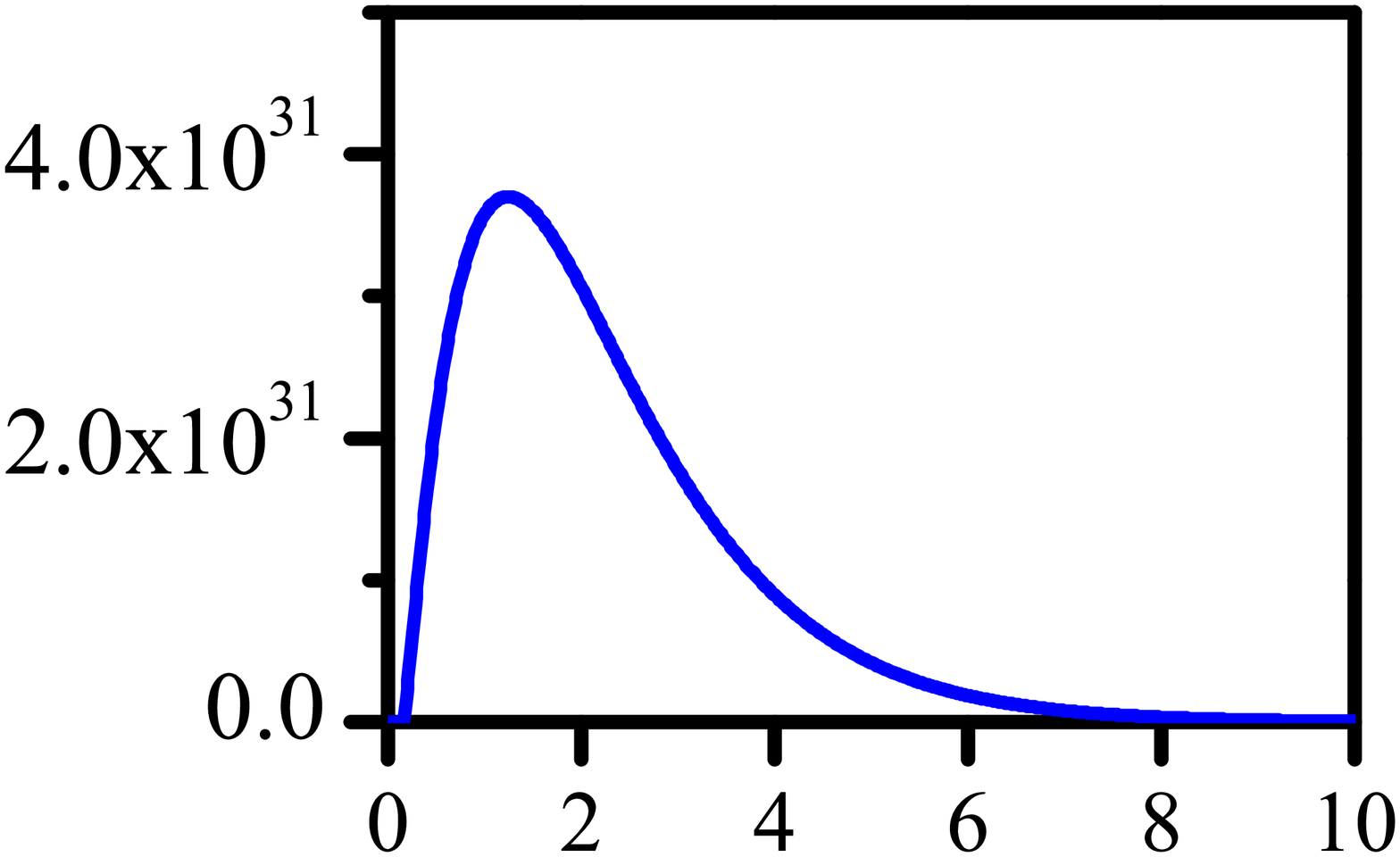}
\vspace{0.8cm}
\caption{(Color online) Comparisons of energy-dependent
evolutions of $|\lambda|$ caused by the competition between all
three sorts of impurities and chemical potential for several initial values of several
representatively initial values of parameters, i.e. $\alpha(0)=5\times10^{-3}$,
$v(0)=10^{-3}$, $Q=10^{-3}$, $\varphi=\pi$ and $\lambda_0=-10^{-4}$: (a) $\mu^0=10^{-5}$
and (b) $\Delta_I^0=10^{-4}$ with increasing the starting values of impurity strength and
chemical potential, respectively. Without loss of generality, all three types of impurities
are assumed to own the equally starting strengths, namely
$\Delta_{C}=\Delta_{G_{1,3}}=\Delta_{M}=\Delta_{I}$. Inset: the enlarged regions of $\lambda(l)$
for $\Delta^0_M=10^{-6}$, $\mu^0=10^{-5}$ and $\Delta^0_M=10^{-4}$, $\mu^0=10^{-4}$
(the basic results are similar for other initial values of impurity strengths (or $\mu$)
and hence not shown here).}\label{Fig14_CGM-mu-lambda_M4}
\end{figure}

\section{Competition between the impurities and a nonzero chemical potential}\label{Sec_imp-mu}

Based on the analyses in Sec.~\ref{Sec_Discussion} and Sec.~\ref{Sec_imp} for 2D SD systems,
we address that the Cooper instability is greatly promoted by the
chemical potential. Conversely, the presence of impurities is harmful to the
Cooper instability. These imply that the low-energy physics of 2D SD materials would be largely
dependent upon which of these two facets is dominant. It therefore is
interesting to ask whether and how the fate of Cooper instability
would be revised or totally changed under the competition between
the impurities and a nonzero chemical potential.

In order to investigate their competition and answer above question, we assume
that the chemical potential and impurities are present simultaneously. To proceed,
we implement the full effective action~(\ref{Eq_S_eff}) and evaluate all one-loop corrections of Fig.~\ref{Fig1_fermion_propagator_correction}-\ref{Fig4_lambda_correction_imp}
with both nonzero $\mu$ and all three sorts of impurities. After paralleling long
but straightforward calculations as provided in Sec.~\ref{Sec_Discussion}
and Sec.~\ref{Sec_imp}, we derive the corresponding energy-dependent evolutions as
\begin{widetext}
\begin{eqnarray}
\frac{d\alpha}{dl}\!\!\!\!&=&\!\!\!\!\frac{[-(\Delta_C+\Delta_{G_1})\mathcal{E}_0]\alpha}{4\pi^2},\,\,\,
\frac{dv}{dl}\!=\!\frac{[-(\Delta_C+\Delta_M)\mathcal{E}_0]v}{4\pi^2},\,\,\,
\frac{d\Delta_C}{dl}\!=\!\frac{\left[\sum_{I\neq C}\Delta_I
\mathcal{E}_1+\Delta_C(\mu^2\mathcal{E}_4-\mathcal{E}_2)\right]\!\Delta_C}{4\pi^2},\label{Eq_alpha_mu-imp}\\
\frac{d\Delta_{G_1}}{dl}\!\!\!\!
&=&\!\!\!\!\frac{\left[\Delta_C(\mathcal{E}_1-2\mathcal{E}_2+\mu\mathcal{E}_3)
-(\Delta_M+\Delta_{G_3})(\mu \mathcal{E}_3+\mathcal{E}_1)+\Delta_{G_1}
(\mu \mathcal{E}_3-\mu^2 \mathcal{E}_4
-\mathcal{E}_2)\right]\!\Delta_{G_1}}{4\pi^2},\\
\frac{d\Delta_{G_3}}{dl}\!\!\!\!
&=&\!\!\!\!\frac{\left[(\Delta_{G_1}+\Delta_M)(\mathcal{E}_1-\mu^2\mathcal{E}_4)
+\Delta_C(\mathcal{E}_1-2\mathcal{E}_0+\mu^2\mathcal{E}_4)-\Delta_{G_3}
(\mathcal{E}_2+2\mu \mathcal{E}_4)\right]\!\Delta_{G_3}}{4\pi^2},\\
\frac{d\Delta_{M}}{dl}\!\!\!\!
&=&\!\!\!\!\frac{\left[(\Delta_{G_1}\!+\!\Delta_{G_3})(\mathcal{E}_1\!-\!2\mathcal{E}_2
\!-\!\mu^2\mathcal{E}_4\!-\!2\mu \mathcal{E}_3)\!-\!\Delta_C(\mathcal{E}_1\!-\!\mu^2\mathcal{E}_4\!+\!2\mu \mathcal{E}_3)\!-\!\Delta_M(\mathcal{E}_2\!+\!2\mu \mathcal{E}_3)
\!+\!4\lambda(\mathcal{E}_1\!-\!\mathcal{E}_2\!-\!\mu^2\mathcal{E}_4)\right]\!\Delta_M}
{4\pi^2},\\
\frac{d\lambda}{dl}\!\!\!\!&=&\!\!\!\!\frac{\left[-4\pi^2+\lambda\left(4\mathcal{D}_2
-4\sum^5_{i=3}\mathcal{D}_i+\mu^2\mathcal{D}_0\right)-\sum_I\Delta_I \mathcal{E}_2\right]\!\lambda}
{4\pi^2},\,\,\,\frac{d\mu}{dl}\!=\!\mu,\label{Eq_lambda_mu-imp}
\end{eqnarray}
\end{widetext}
with designating two new parameters,
\begin{eqnarray}
\mathcal{E}_3&=&\int^{\frac{\pi}{2}}_{-\frac{\pi}{2}}d\theta
\frac{\alpha\cos\theta}{\sqrt{\cos\theta}
(\alpha^2\cos^2\theta+v^2\sin^2\theta)^2}~\label{Eq_E_3},\\
\mathcal{E}_4&=&\int^{\frac{\pi}{2}}_{-\frac{\pi}{2}}d\theta
\frac{1}{\sqrt{\cos\theta}(\alpha^2\cos^2\theta+v^2\sin^2\theta)^2}~\label{Eq_E_4}.
\end{eqnarray}

In order to compare with the $\mu=0$ situation discussed in Sec.~\ref{Sec_imp}, we employ
the same initial values of related interaction parameters used in Fig.~\ref{Fig12_Mass_contrast-lambda-M4}
and Fig.~\ref{Fig13_CGM_contrast-lambda-M4}. Based on the numerical calculations of these
coupled equations~(\ref{Eq_alpha_mu-imp})-(\ref{Eq_lambda_mu-imp}), the primary results
are delineated in Fig.~\ref{Fig14_CGM-mu-lambda_M4}, which clearly shows the
evolutions of fermion-fermion strengths for the presence of distinct values of
the chemical potential and impurities. These results exhibit several interesting
features. At first, once the initial value of chemical potential is small, for instance
$\mu^0=10^{-5}$, the impurities dominate over chemical potential as
shown in Fig.~\ref{Fig14_CGM-mu-lambda_M4}(a). In this respect, the fate of Cooper
instability is governed by the impurities. However, compared to the
$\mu=0$ case in Fig.~\ref{Fig13_CGM_contrast-lambda-M4}, the chemical potential
slightly enhance the saturated peak of the fermion-fermion strength. Next,
as displayed in Fig.~\ref{Fig14_CGM-mu-lambda_M4}(b), we can manifestly find
that the chemical potential becomes prevailing while its starting value is sufficient large.
In short, either impurity or chemical potential is able to play a vital role
in pinning down the fate of the Cooper instability at the low-energy regime.
In addition, whose effect is dominant, to a large extent, relies upon their
initial values and the intimate competition between them.

\section{Summary}\label{Sec_summary}

In summary, stimulated by the even more unconventional features of 2D SD compared to the DSM materials,
we primarily investigate whether and how the Copper instability that is associated with the superconductivity
can be induced by an attractive Cooper-pairing interaction in the 2D SD semimetals as well as influenced by
the impurity scatterings at zero chemical potential. In addition, the effects of a finite chemical potential
at clean-limit are carefully studied. Moreover, how the competition between
the impurities and a finite chemical potential influence the Cooper instability is also
briefly examined.

Concretely, we introduce the Cooper-pairing interaction stemmed from an attractive
fermion-fermion interaction and fermion-impurity interaction obtained via averaging
impurity potential to build our effective field theory. In order to take into account
these distinct sorts of physical degrees of freedoms on the same footing, we adopt
the momentum-shell RG approach. Upon carrying out the standard RG analysis,
we collect the one-loop corrections due to
the Cooper-pairing and fermion-impurity interactions and next derive the energy-dependent
evolutions of interaction parameters at both $\mu=0$ and $\mu\neq0$. To proceed, we employ
these RG flows to attentively examine the emergence of the Cooper instability in the low-energy
regime. Taking $\mu=0$ at first, we find that the Cooper-pairing strength $\lambda$ evolves
towards zero upon lowering energy scales at the presence of only tree-level corrections,
namely Cooper instability cannot be activated. After including the
one-loop corrections, we find the BCS subchannel correction of
Cooper-pairing interaction vanishes and the RG running of parameter $\lambda$
only depends upon the correction from the summation of ZS and $\mathrm{ZS}'$ subchannels
while the internal transfer momentum $\mathbf{Q}$ is nonzero. This is sharply contrast to the DSM systems,
at which the BCS subchannlel contributes dominantly to the parameter $\lambda$ at $\mu=0$.
After performing both analytical and numerical analyses, we conclude that
the summation of ZS and $\mathrm{ZS}'$ contributions, which are dependent
upon the strength and direction of the transfer momentum $\mathbf{Q}$,
is crucial to the emergence of Cooper instability. Under certain
circumstance, the Cooper instability can be triggered once the strength and direction of
$\mathbf{Q}$ are reasonable and the initial strength of $|\lambda(0)|$ exceeds some critical value.
Additionally, we move to the $\mu\neq0$ situation. The RG analysis tells us that the
parameter $\mu$ is a relevant quantity in term of the RG language. It directly
suggests that the Cooper theorem~\cite{BCS1957PR} would be restored. In other words,
any weak Cooper-pairing interaction can induce the Cooper instability and thus
a $\mu$-tuned phase transition is expected. Moreover, we carefully study
how three primary types of impurities at zero chemical potential impact the
Cooper instability. For completeness, the influence of
competition between the impurities and a finite chemical
potential on Cooper instability is also briefly investigated. In short,
we find that which of facets among three
types of impurities and a finite chemical potential is dominant
largely hinges upon their initial values
and the competition between them is of remarkable significance to
determine the final fate of Cooper instability at the low-energy regime.

Studying the superconductivity in kinds of semimetals is an intriguing clue to
reveal the microscopic mechanism of unconventional superconductors, for instance
the cuprate high-$T_c$ materials~\cite{Lee2006RMP}, iron-based compounds~\cite{Chubukov2012ARCMP,Chubukov2017RPP},
layered organic~\cite{Powell2011RPP} and heavy-fermion superconductors~\cite{Stewart1984RMP,Steglich2016RPP}.
It is particularly worth mentioning that the Mott insulator and superconductor have been
realized very recently in the twisted bilayer graphene~\cite{Herrero2018Nature1,Herrero2018Nature2}.
We therefore wish our study would be helpful to uncover the unique features of 2D SD materials
and explore their relations with the superconductors.

\section*{ACKNOWLEDGEMENTS}

J.W. is supported by the National Natural Science Foundation of China
under Grant No. 11504360. We acknowledge Prof. W. Liu for useful discussions.

\appendix

\section{Related coefficients}\label{Appendix_coefficients}

Here we gather the related coefficients introduced in Sec~\ref{RG_clean_case}
and Sec.~\ref{Sec_Discussion} for the discussions of clean-limit case as follows,
\begin{widetext}
\begin{eqnarray}
\mathcal{D}_0
&\equiv&\int^{\frac{\pi}{2}}_{\frac{-\pi}{2}}d\theta\frac{1}
{{(\alpha^2\cos^2\theta+v^2\sin^2\theta)}^{\frac{3}{2}}\sqrt{\cos\theta}},\hspace{0.16cm}
\mathcal{D}_1\equiv\int^{\frac{\pi}{2}}_{\frac{-\pi}{2}}d\theta
\frac{\alpha^2\cos^2\theta}{{(\alpha^2\cos^2\theta+v^2\sin^2\theta)}^{\frac{3}{2}}\sqrt{\cos\theta}}
,\label{Eq_D_1}
\\
\mathcal{D}_2&\equiv&\int^{\frac{\pi}{2}}_{-\frac{\pi}{2}}d\theta
\frac{\alpha^2Q\cos\varphi\cos^{\frac{3}{2}}\theta}
{(\alpha^2\cos^2\theta+v^2\sin^2\theta)^{\frac{3}{2}}\sqrt{\cos\theta}},\hspace{0.16cm}
\mathcal{D}_3\equiv\int^{\frac{\pi}{2}}_{-\frac{\pi}{2}}d\theta\frac{[6Q\alpha^2\cos^{\frac{3}{2}}
\theta\cos\varphi(v^2\sin^2\theta+\alpha^2Q\cos\varphi\cos^{\frac{3}{2}}\theta)]}
{(\alpha^2\cos^2\theta+v^2\sin^2\theta)^{\frac{5}{2}}\sqrt{\cos\theta}},
\\
\mathcal{D}_4&\equiv&\int^{\frac{\pi}{2}}_{-\frac{\pi}{2}}d\theta
\frac{[3\alpha^2Q\cos^{\frac{3}{2}}\theta\cos\varphi(v^2\sin^2\theta+\alpha^2Q\cos\varphi
\cos^\frac{3}{2}\theta-15Q^2\sin^2\varphi v^4\sin^2\theta)]}
{(\alpha^2\cos^2\theta+v^2\sin^2\theta)^{\frac{7}{2}}\sqrt{\cos\theta}},
\\
\mathcal{D}_5&\equiv&\int^{\frac{\pi}{2}}_{-\frac{\pi}{2}}d\theta
\frac{[60\alpha^6Q^3\cos^3\varphi\cos^{\frac{9}{2}}
\theta( v^2\sin^2\theta+\alpha^2Q\cos\varphi\cos^\frac{3}{2}
\theta)]}{(\alpha^2\cos^2\theta+v^2\sin^2\theta)^{\frac{9}{2}}\sqrt{\cos\theta}},\label{Eq_D_5}\\
\mathcal{M}&\equiv&Q^4\int^{\frac{\pi}{2}}_{-\frac{\pi}{2}}d\theta
\frac{(60\alpha^8\cos^{6}\theta)\cos^4\varphi}{(\alpha^2\cos^2\theta
+v^2\sin^2\theta)^{\frac{9}{2}}\sqrt{\cos\theta}},\label{Eq-coeff_M}\\
\mathcal{N}
&\equiv&Q^3\int^{\frac{\pi}{2}}_{-\frac{\pi}{2}}d\theta\frac{[60\alpha^6\cos^{\frac{9}{2}}
\theta v^2\sin^2\theta\cos^3\varphi+45\alpha^2v^4 \cos^{\frac{3}{2}}\theta
\sin^2\theta(\alpha^2\cos^2\theta+v^2\sin^2\theta)]\cos^3\varphi}
{(\alpha^2\cos^2\theta+v^2\sin^2\theta)^{\frac{9}{2}}
\sqrt{\cos\theta}},\label{Eq-coeff_N}\\
\mathcal{O}
&\equiv&Q^2\int^{\frac{\pi}{2}}_{-\frac{\pi}{2}}d\theta\frac{[6\alpha^4\cos^{3}
\theta(\alpha^2\cos^2\theta+v^2\sin^2\theta)+
3\alpha^4\cos^{3}
\theta(\alpha^2\cos^2\theta+v^2\sin^2\theta)^2]\cos^2\varphi
}{(\alpha^2\cos^2\theta+v^2\sin^2\theta)^{\frac{9}{2}}
\sqrt{\cos\theta}},\label{Eq-coeff_O}\\
\mathcal{P}
&\equiv&Q\int^{\frac{\pi}{2}}_{-\frac{\pi}{2}}d\theta
\frac{1}{(\alpha^2\cos^2\theta+v^2\sin^2\theta)^{\frac{9}{2}}\sqrt{\cos\theta}}
[3\alpha^2\cos^{\frac{3}{2}}\theta v^2\sin^2\theta
(\alpha^2\cos^2\theta+v^2\sin^2\theta)\nonumber\\
&&+\alpha^2 \cos^{\frac{3}{2}}\theta (3\alpha^4v^2\sin^2\theta\cos^4\theta+12\alpha^2v^4\cos^2\theta
\sin^4\theta-\alpha^6\cos^6\theta+
5v^6\sin^6\theta)\nonumber\\
&&-45\alpha^2v^4 Q^2\cos^{\frac{3}{2}}\theta
\sin^2\theta(\alpha^2\cos^2\theta+v^2\sin^2\theta)]\cos\varphi,\label{Eq-coeff_P}.
\end{eqnarray}
\end{widetext}

\section{One-loop calculations for the presence
of impurities at $\mu=0$}\label{Appendix_impurity-calculations}

In order to capture the effects of impurities on Cooper instability, we adopt the effective
action~(\ref{Eq_S_eff}) by assuming $\mu=0$. This indicates
that several additional one-loop Feynman diagrams are involved as illuminated in
Figs.~\ref{Fig1_fermion_propagator_correction}, \ref{Fig2_fermion_propagator_correction_imp}
and \ref{Fig4_lambda_correction_imp} owing to the impurity scatterings. The evaluations of these one-loop corrections are tedious but straightforward~\cite{Wang2011PRB,Wang2013PRB,Wang2017PRB_BCS}.

To be specific, carrying out the analogous analyses in Sec.~\ref{RG_clean_case} gives rise to
the one-loop corrections from fermion-impurity as follows.
Fig.~\ref{Fig1_fermion_propagator_correction} gives rise to
\begin{eqnarray}
\delta\Sigma_{\mathcal{I}}
\!\!&=&\!\!\frac{-\sum_I\Delta_I\mathcal{E}_0}{8\pi^2}(-i\omega)l\nonumber\\
\!\!\!\!\!\!&&+\frac{(\Delta_C+\Delta_{G1}
-\Delta_M-\Delta_{G_3})\mathcal{E}_0}{8\pi^2}(\alpha k^2_x \sigma_1)l\nonumber\\
\!\!\!\!\!\!&&+\frac{(\Delta_C+\Delta_M
-\Delta_{G1}-\Delta_{G_3})\mathcal{E}_0}{8\pi^2}(v k_y \sigma_2)l,
\end{eqnarray}
where $C$, $G_{1,3}$, and $M$ defined in Sec.~\ref{Subsec_imp}
correspond to the random chemical potential, random gauge potential and random mass, respectively.
Additionally, the coefficient $\mathcal{E}_0$ is defined as
\begin{eqnarray}
\mathcal{E}_0
&\equiv&\int^{\frac{\pi}{2}}_{-\frac{\pi}{2}}d\theta
\frac{1}{\sqrt{\cos\theta}(\alpha^2 \cos^2\theta+v^2\sin^2\theta)}\label{Eq_E_0}.
\end{eqnarray}
The strength of fermion-impurity coupling provided in
Fig.~\ref{Fig2_fermion_propagator_correction_imp}. Summing up all five subfigures
yields to
\begin{eqnarray}
\delta\Delta_C
&=&\frac{1}{4\pi^2}\Bigl[\Delta_C\mathcal{E}_1+\sum_{I\neq C}\Delta_I(\mathcal{E}_0+\mathcal{E}_1)\Bigr],\\
\delta\Delta_{G_1}
&=&\frac{1}{4\pi^2}\Bigl[(\Delta_M+\Delta_{G_3})\mathcal{E}_2+\Delta_{G_1} \mathcal{E}_1\nonumber\\
&&+\Delta_C (2\mathcal{E}_1-\mathcal{E}_2)\Bigr],\\
\delta\Delta_{M}
&=&\frac{1}{4\pi^2}\Bigl[(\Delta_{G_1}+\Delta_{G_3})(2\mathcal{E}_1-\mathcal{E}_2)
+\Delta_C\mathcal{E}_2\nonumber\\
&&+\Delta_M\mathcal{E}_1+4\lambda(\mathcal{E}_1-\mathcal{E}_2)\Bigr],\\
\delta\Delta_{G_3}
&=&\frac{1}{4\pi^2}\Bigl[(\Delta_M+\Delta_{G_1})(\mathcal{E}_0+\mathcal{E}_1)\nonumber\\
&&+\Delta_C(\mathcal{E}_1-\mathcal{E}_0)+\Delta_{G_3}\mathcal{E}_1\Bigr],
\end{eqnarray}
where $\mathcal{E}_1$ and $\mathcal{E}_2$ are respectively nominated as
\begin{eqnarray}
\mathcal{E}_1&\equiv&\int^\frac{\pi}{2}_{-\frac{\pi}{2}}d\theta\frac{\alpha^2 \cos^2\theta}
{\sqrt{\cos\theta}(\alpha^2 \cos^2\theta+v^2 \sin^2\theta)^2},\label{Eq_E_1}\\
\mathcal{E}_2&\equiv&\int^\frac{\pi}{2}_{-\frac{\pi}{2}}d\theta
\frac{v^2\sin^2\theta}{\sqrt{\cos\theta}(\alpha^2 \cos^2\theta+v^2\sin^2\theta)^2}.\label{Eq_E_2}
\end{eqnarray}
Finally, one-loop corrections to Cooper coupling $\lambda$
depicted in Fig~\ref{Fig4_lambda_correction_imp} are left with
\begin{eqnarray}
\delta \lambda_{\mathcal{I}}=
\frac{\sum_{I}\Delta_I\mathcal{E}_1}{4\pi^2}.\label{Eq_Self_energy_0}
\end{eqnarray}
Before going further, it is of very importance to highlight the main differences between
2D DSM~\cite{Sondhi2013PRB,Sondhi2014PRB,Wang2017PRB_BCS} and 2D SD materials~\cite{Hasegawa2006PRB,Pardo2009PRL,Katayama2006JPSJ,
Dietl2008PRL,Delplace2010PRB,Banerjee2009PRL,Banerjee2012PRB,Wu2014Expre,Saha2016PRB,
Uchoa2017PRB,Yang-Isobe2014NP,Isobe2016PRL,
Cho-Moon2016SR,WLZ2017PRB,Roy2018PRX,Wang2018JPCM}. For the 2D DSM systems, one
can realize that one-loop corrections by impurity scatterings, namely
Fig~\ref{Fig2_fermion_propagator_correction_imp}(ii)-(v)
vanish due to the linear dispersions for both $k_x$ and $k_y$ directions.
In addition, Fig.~\ref{Fig4_lambda_correction_imp}(i) is also neutralized by
Fig.~\ref{Fig4_lambda_correction_imp}(ii).
In a sharp contrast, they contribute very nonzero
values for our 2D SD systems, which significantly
modify the evolution of parameter $\lambda$. According to above one-loop corrections, we find the fermionic
field gains a nonzero anomalous fermion dimension~\cite{Nersesyan1995NPB,Huh2008PRB,Wang2011PRB}, namely
\begin{eqnarray}
\eta=\frac{\sum_I\Delta_I \mathcal{E}_0}{16\pi^2}.
\end{eqnarray}


\end{CJK*}

\end{document}